\documentclass[aps,pra, twocolumn,notitlepage,groupedaddress,reprint]{revtex4-1}
\usepackage{pifont}%
\usepackage{bbold}%
\usepackage{braket}
\usepackage[dvipsnames]{xcolor} %
\usepackage{tikz}
\usepackage{qcircuit}
\usepackage{amsmath, amssymb, graphicx, bm}
\usepackage{comment}
\usepackage{pifont}%
\usepackage{pythonhighlight}
\usepackage{hyperref}
\usepackage{algorithm}
\usepackage{algpseudocode}
\algrenewcommand\algorithmicrequire{\textbf{Input:}}
\algrenewcommand\algorithmicensure{\textbf{Output:}}

\usepackage{makecell}

\usepackage{tensor}
\usepackage{amsthm}
\usepackage{amsfonts}
\usepackage{mathrsfs}
\usepackage{mathtools}
\usepackage{MnSymbol}
\usepackage{tikz-cd}
\usetikzlibrary{decorations.pathmorphing}

\usepackage{tikz}
\usetikzlibrary{shapes.geometric, positioning, arrows.meta}

\theoremstyle{plain}

\DeclareMathAlphabet{\mathpzc}{OT1}{pzc}{m}{it}

\begin{document}
\title{Optimal local oscillators for the homodyne detection of multiphoton states}
\author{Gisell Lorena Osorio}
\email{gisell-lorena.osorio-osorio@polymtl.ca}
\affiliation{Department of Engineering Physics, \'Ecole polytechnique de Montr\'eal, Montr\'eal, QC, H3T 1J4, Canada}
\author{Paul Virally}
\affiliation{Department of Engineering Physics, \'Ecole polytechnique de Montr\'eal, Montr\'eal, QC, H3T 1J4, Canada}
\author{Sean Molesky}
\affiliation{Department of Engineering Physics, \'Ecole polytechnique de Montr\'eal, Montr\'eal, QC, H3T 1J4, Canada}
\author{Nicol\'as Quesada}
\affiliation{Department of Engineering Physics, \'Ecole polytechnique de Montr\'eal, Montr\'eal, QC, H3T 1J4, Canada}

\begin{abstract}  

      We propose a framework for optimizing pulsed local oscillators (LO) for homodyne detection of multiphoton-number states by exploiting the tensor structure of their joint-spectral amplitude (JSA). We show that finding the optimal LO is equivalent to computing the JSA tensor's leading unitary eigenpair and that the factor matrices of the JSA's Tucker higher-order singular value decomposition (HOSVD) coincide with the Schmidt modes of the photon number state's single-particle reduced density matrix. We use the HOSVD to bound the optimal homodyne visibility and to initialize gradient based optimization. In simulated JSAs, weakly correlated, few-mode states reach near-unity visibility, with the leading HOSVD mode being the optimal LO, while strongly correlated, multimode states require full optimization and saturate below unit visibility even at the true optimum. These results offer a practical route to design LOs for homodyne detection experiments with realistic sources of photon-number states, which contributes to the practical implementation of sources of non-Gaussian quantum states.
\end{abstract}

\maketitle

\section{Introduction}

To harness the potential of quantum technologies, rigorous tools are needed to characterize the behavior of quantum states and devices. 
Quantum tomography provides the theoretical and experimental framework for state reconstruction, and has become a cornerstone of quantum information science\;\cite{deGois2024,mele2025learning,wang2025quantum}. Among the experimental techniques used to perform quantum tomography, homodyne detection stands out as one of the most widely used in quantum optics, due to its ability to access the quadrature statistics of a quantum field with high precision \;\cite{lvovsky2009continuous,tiunov2020experimental,fedotova2023continuous}.

In homodyne detection, the local oscillator (LO) establishes the reference phase and signal components that are actually measured. If the LO mode differs from that of the signal containing relevant features in any relevant degrees of freedom (polarization, spatial profile, or spectral distribution), part of that information is inevitably lost due to this mismatch. Therefore, appropriately designing the LO is essential for revealing a complete picture of the quadrature statistics of the signal of interest\;\cite{lvovsky2009continuous,fabre2020modes, shapiro1997optimizing}. This becomes particularly significant when attempting to identify non-classical signatures of quantum states or when attempting to measure non-Gaussian states in quantum tomography experiments\;\cite{ kawasaki2024broadband, osorio2025strategies}.

Photon-number states, which are intrinsically non-Gaussian, present a direct motivation for this problem for both fundamental and practical reasons. They exhibit a Wigner distribution with negativity and a sharp oscillatory structure, and the ability to resolve features is highly sensitive to the choice of the measurement mode\;\cite{lvovsky2001quantum, israel2012experimental}. 
These states are also becoming increasingly relevant experimentally: while heralded single photons and photon pairs from \textit{spontaneous parametric downconversion} (PDC) and \textit{spontaneous four-wave mixing} (FWM) have long been the standard nonclassical resources, recent advances in nonlinear photonics, ranging from  third-order parametric downconversion\;\cite{mougin2026photon, banic2022resonant,osorio2025strategies,okoth2019seeded}, cascaded PDC and FWM in integrated platforms\;\cite{krapick2016chip,fontaine2026photon}, and materials with large high-order susceptibilities\;\cite{skachkov2026chip}, are opening direct access to higher-order multiphoton states\;\cite{willemann2025can, tiedau2019scalability}. These states are central to applications in quantum metrology, quantum information processing and quantum computing protocols which rely on non-Gaussian resources\;\cite{lloyd1999quantum, weedbrook2012gaussian, Walschaers2021nonGaussian, grun2022protocol, deng2024quantum}. 

Here, we address the problem of optimizing the local oscillator in homodyne detection schemes to measure the quadratures of photon-number states. In the continuous-wave regime, such states typically exhibit a multimode spectral structure, which complicates their characterization. Ideally, one would engineer the source to produce single-mode photon-number states\;\cite{osorio2025strategies}; however, this remains experimentally challenging. Instead, we take a complementary approach: assuming ideal spatial mode-matching between the local oscillator and the signal field, we focus on tailoring the spectral distribution of the local oscillator to maximize detection efficiency for multimode photon-number states. This multimode spectral structure is captured by the joint spectral amplitude (JSA) of the state, which for an $N-$photon state is an $N-$order tensor\;\cite{perarnau2020multimode}, whose geometry is critical for the optimization of the detection. As the order of the JSA grows, its dimensionality renders the optimization landscape more complex, calling for new design strategies. To approach this problem, we examine the tensor structure of the multi-photon JSA and show that optimizing the spectral distribution of the LO is equivalent to a tensor eigenpair problem on the JSA. Furthermore, we propose the Tucker higher-order singular value decomposition (HOSVD) as a useful spectral analysis tool for high-dimensional quantum states\;\cite{bell2010higher,stanislav2023faster}, whose wavefunctions are intrinsically tensorial. This is complementary to other approaches, such as tensor networks and matrix product states\;\cite{florido2024product}. 

Using this framework, we derive analytical bounds for the maximum achievable visibility in homodyne detection based on the HOSVD of the JSA, and we find that for nearly separable states the HOSVD directly yields the optimal LO without any further optimization. For the general, non-separable case, we introduce a gradient-based algorithm informed by the HOSVD and validate it on two families of JSAs, Gaussian and those arising from high-order parametric downconversion, across photon numbers $N\,\in\,\{3,4,5,6\}$ and a range of spectral correlation degrees. Our algorithm combines two complementary approaches: gradient descent and homotopy continuation\;\cite{chen2016computing}. The flexibility and efficiency of gradient-based methods are combined with a topology optimization strategy which takes into account the geometry of the JSA tensor. We find that our algorithm reliably converges to the global optimum across all tested cases, and that it scales favorably compared to other optimization methods to find tensor eigenpairs. Our strategy is oriented toward realistic experimental devices, and contributes to broader efforts aimed not only at improving non-Gaussian quantum state sources, but also at optimizing their detection, which is a critical step toward their practical usability.

This paper is organized as follows. In Section \ref{sec:quadrature} we review the quadrature statistics of photon number states and their relationship to the JSA. In Section \ref{sec:optim_problem} we formulate the optimal LO problem and establish its equivalence to the tensor eigenpair problem. In Section \ref{sec:eigen_compute} we discuss the computation of tensor eigenpairs and motivate the use of gradient-based methods. In Section \ref{sec:algorithm} we present our algorithm based on the HOSVD, and include an analytical derivation of bounds for the maximum efficiency of the homodyne detection of photon-number states. Finally, in Section \ref{sec:results} we present numerical experiments, and in Section \ref{sec:discussion} we discuss the results and outline directions for future work.

\section{Theoretical framework}

\subsection{\label{sec:quadrature}Quadrature distribution of photon number states}

Consider the $N-$photon state: 
\begin{align}\label{eq:Psi_1}
    \ket{\Psi_N}=\frac{1}{\sqrt{N!}} \int d\omega_1...d\omega_N ~ \psi (\omega_1,...,\omega_N) \nonumber \\  a^\dagger(\omega_1)...a^\dagger(\omega_N) \ket{\text{vac}},
\end{align}
where $\ket{\text{vac}}$ is the vacuum state, $\psi (\omega_1,...,\omega_N)$ is the $N-$photon wavefunction or joint-spectral amplitude (JSA) and $a^\dagger(\omega_i)$ are creation operators that satisfy the commutation relations
\begin{align}    [a(\omega_i),a^\dagger(\omega_j)]=\delta(\omega_i-\omega_j), \quad [a(\omega_i),a(\omega_j)]=0. 
\end{align}
$\psi$ is invariant under permutations of the frequency variables and is normalized such that
\begin{align}
    \int d\omega_1...d\omega_N  |\psi (\omega_1,...,\omega_N)|^2=1.
\end{align}

We define generalized mode creation operators\;\cite{rohde2007spectral}
\begin{align}\label{eq:broadband_operator}
 A_i^\dagger=\int d\omega_j f_i(\omega_j) a^\dagger(\omega_j),   
\end{align}
where the spectral mode functions $\{f_i(\omega)\}_{i=0}^\infty$ form a complete and orthonormal basis, satisfying
\begin{equation}\label{eq:complete_orthonorm}
    \sum_{i=0}^\infty f_i (\omega) f_i^*(\omega')=\delta(\omega-\omega'), \quad 
    \int d\omega f_i^*(\omega)f_j(\omega)=\delta_{ij}.
\end{equation}
We assume that the field operators and temporal modes are centered at optical carrier frequencies $\omega_{i0}>0$ with bandwidths $\Delta \omega_i \ll \omega_{i0}$. Under these assumptions, the integration limits in Eqs. \eqref{eq:Psi_1} to \eqref{eq:complete_orthonorm} may be defined from $-\infty$ to $+\infty$. Using Eqs. \eqref{eq:broadband_operator}-\eqref{eq:complete_orthonorm}, we can write the $N-$photon state (Eq. \eqref{eq:Psi_1}) in terms of the broadband operators as
\begin{align}\label{eq:Psi_2}
    \ket{\Psi_N}
=\frac{1}{\sqrt{N!}} \int d\omega_1...d\omega_N ~ \psi (\omega_1,...,\omega_N) \nonumber \\ \left[ \prod _j^N \sum_{i=0}^\infty f^\ast_i(\omega_j)A_i^\dagger \right]\ket{\text{vac}}. 
\end{align}

In a homodyne detection experiment, the measurement outcome corresponds to the quadrature $x_\theta$ in the mode defined by the local oscillator, and any orthogonal modes do not contribute to the homodyne signal. Let the local oscillator mode be described by a normalized mode function $f_0(\omega)$ and let us introduce a broadband generalized quadrature operator
\begin{equation}
    x_\theta=\frac{A_0^\dagger e^{i\theta}+A_0 e^{-i\theta}}{2},
\end{equation}
where $A_0^\dagger=\int d\omega f_0(\omega)a^\dagger(\omega)$. We choose $f_0(\omega)$ to be an element of the discrete basis of orthonormal functions $\{f_i(\omega)\}_{i=0}^\infty$ introduced previously. The quadrature distribution for the $N-$photon state is $p(x|\theta)=|\braket{x_\theta|\Psi_N}|^2$, where
\begin{align}    \braket{x_\theta|\Psi_N}&=\eta  \bra{x_\theta} \frac{A_0^{\dagger N}}{\sqrt{N!}} \ket{\text{vac}} \\
    &=\eta \frac{H_N(x)}{\sqrt{2^N N! \sqrt{\pi}}} \exp{\left(-\tfrac{1}{2} x^2\right)} \exp{\left(-i N \theta\right)}.
\end{align}

$H_N(x)$ denotes the Hermite polynomials\;\cite{barnett1997methods} and
\begin{equation}\label{eq:eta}
    \eta=\int d\omega_1...d\omega_N ~ \psi (\omega_1,...,\omega_N) \prod _j^N f^\ast_0(\omega_j).
\end{equation}
The quantity $\eta$ is assumed real without loss of generality. The function $f_0(\omega)$ represents the spectral distribution of the local oscillator. Maximizing the overlap $\eta$ is required to maximize the detection efficiency of photon-number states. This becomes especially important under realistic experimental conditions, where the prepared state typically contains a strong vacuum contribution, $\ket{\Phi}\approx \ket{\text{vac}}+\epsilon\ket{\Psi_N}$ with $\epsilon\ll 1$. In such a case, maximizing $\eta$ is essential to enable the detection of any non-Gaussian deviations of photon-number states in their quadrature statistics. A suboptimal LO can suppress the contribution of the $\ket{\Psi_N}$ component, potentially rendering it undetectable. Thus, maximizing $\eta$ is not merely a matter of efficiency but a prerequisite for revealing the non-classical character of the state. Since the JSA is fixed by the properties of the source, the goal is to find a local oscillator distribution that maximizes $\eta$ for that particular JSA.

\subsection{\label{sec:optim_problem} Optimal local oscillators as a tensor eigenvalue problem}

In practice, for numerical implementations, we discretize the frequency domain using a uniform grid. We define  $f_i\equiv (\Delta \omega)^{1/2} f_0(\omega_i)$, so that the continuous normalization $\int d\omega_i~ |f_0(\omega_i)|^2=1$ becomes the discrete condition $\sum_i |f_i|^2=1$ in the limit that $\Delta \omega \xrightarrow{}0$. Similarly, the discretized JSA is defined as $\psi_{i_{\scriptscriptstyle 1}...i_{\scriptscriptstyle N}}\equiv (\Delta \omega)^{N/2} \psi (\omega_1,...,\omega_N) $, with normalization $\sum_{i_{\scriptscriptstyle 1}...i_{\scriptscriptstyle N}}|\psi_{i_{\scriptscriptstyle 1}...i_{\scriptscriptstyle N}}|^2=1$. The functional $\eta$ (eq. \eqref{eq:eta}) can then be written as
\begin{equation} \label{eq:discrete_eta}
   \eta= \bm{\psi} \cdot \bm{f}^{\otimes N}:= \sum_{i_{\scriptscriptstyle 1},i_{\scriptscriptstyle 2},...,i_{\scriptscriptstyle N}} \psi_{i_{\scriptscriptstyle 1}i_{\scriptscriptstyle 2}...i_{\scriptscriptstyle N}} f^\ast_{i_{\scriptscriptstyle 1}} f^\ast_{i_{\scriptscriptstyle 2}}...f^\ast_{i_{\scriptscriptstyle N}}.
\end{equation}
Here, $\bm{f}^{\otimes N}:=\bm{f} \otimes\bm{f} \otimes...\otimes \bm{f}$ ($N$ times) is the $N-$th tensor power of $\bm{f}$, with components $\left(\bm{f}^{\otimes N}\right)_{i_{\scriptscriptstyle 1}i_{\scriptscriptstyle 2}...i_{\scriptscriptstyle N}}=f_{i_{\scriptscriptstyle 1}} f_{i_{\scriptscriptstyle 2}}...f_{i_{\scriptscriptstyle N}}$, and the notation $\bm{\psi} \cdot \bm{f}^{\otimes N}$ denotes the inner product between $\bm{f}^{\otimes N}$ and the tensor $\bm{\psi}$, equivalent to $\braket{\bm{f}^{\otimes N},\bm{\psi}}$, where 
\begin{equation} \label{eq:inner_product}
\braket{A,B}=\sum_{j_{\scriptscriptstyle 1},...,j_{\scriptscriptstyle m}} A^\ast_{j_{\scriptscriptstyle 1},...,j_{\scriptscriptstyle m}} B_{j_{\scriptscriptstyle 1},...,j_{\scriptscriptstyle m}}    
\end{equation}
denotes the inner product and  $A^\ast_{j_{\scriptscriptstyle 1},...,j_{\scriptscriptstyle m}} $ is the complex conjugate of $A_{j_{\scriptscriptstyle 1},...,j_{\scriptscriptstyle m}} $. The tensor $\bm{\psi}$ is totally symmetric \footnote{Also called \textit{supersymmetric} in the mathematics literature.}, meaning that it remains invariant under all permutations of the indices. This symmetry in the $N-$photon JSA is a natural consequence of the bosonic nature of photon number states, arising from the indistinguishability of identical particles. 

The problem of finding the optimal local oscillator for photon number states is expressed as the nonlinear optimization problem
\begin{equation}\label{eq:optim_prob}
\begin{aligned}
\max_{\bm{f}} \;  \bm{\psi} \cdot \bm{f}^{\otimes N} 
\quad \text{subject to} \quad \bm{f}^* \bm{f} = 1.
\end{aligned}
\end{equation}

Writing the normalization constraint as $c(\bm{f})=\left(\bm{f}^* \bm{f} -1\right)=0$, its associated Lagrangian is 
\begin{align} \label{eq:lagrangian}
\mathcal{L}(\bm{f},\alpha)=\bm{\psi} \cdot \bm{f}^{\otimes N}+\alpha ~c(\bm{f})
\end{align}
Its first derivative with respect to $\bm{f}^\ast$ is
\begin{align}\label{eq:grad_lagrangian}
\nabla \mathcal{L}(\bm{f},\alpha)=N\bm{\psi} \cdot \bm{f}^{\otimes(N-1)}+\alpha\bm{f},
\end{align}
where
\begin{equation} \label{eq:tensor_contraction}
   \left(\bm{\psi} \cdot \bm{f}^{\otimes(N-1)} \right)_{i_{\scriptscriptstyle 1}}= \sum_{i_{\scriptscriptstyle 2},...,i_{\scriptscriptstyle N}} \psi_{i_{\scriptscriptstyle 1}i_{\scriptscriptstyle 2}...i_{\scriptscriptstyle N}}  f^\ast_{i_{\scriptscriptstyle 2}}...f^\ast_{i_{\scriptscriptstyle N}}.
\end{equation}
Any pair $(\bm{\tilde{f}},\tilde{\alpha})$ satisfying the Karush-Kuhn-Tucker (KKT) conditions $\nabla \mathcal{L}(\bm{\tilde{f}},\tilde{\alpha})=0$ and $c(\bm{\tilde{f}})=0$ is a constrained stationary point\;\cite{kolda2011shifted}. Furthermore, by setting Eq. \eqref{eq:grad_lagrangian} equal to zero we have
\begin{align}\label{eq:eigenvalue}
\bm{\psi} \cdot \bm{f}^{\otimes(N-1)}=\lambda\bm{f},
\end{align}
with $\lambda=-\alpha/N$, which precisely coincides with the complex generalization of the $L^2-eigenpairs$,  originally introduced by Qi\;\cite{qi2005eigenvalues,qi2007eigenvalues} and  Lim\;\cite{lim2005singular} for real-valued symmetric tensors \footnote{Real eigenpairs of real-valued tensors are commonly referred to as \textit{z-eigenvalues} in the literature.} and subsequently extended to complex symmetric tensors, as studied in\;\cite{muller2022robust,che2020theory, zhang2020iterative, ni2014geometric}.  The resulting eigenpairs are also referred to in the literature as \textit{unitary eigenpairs} or \textit{u-eigenpairs}. 

The case of $N=2$ can be easily solved by doing a Takagi-Autonne decomposition\;\cite{houde2024matrix} 
\begin{align}
    \psi_{ij}=\sum_k r_k ~g_{ki} g_{kj},
\end{align}
where $r_i\geq r_{i+1}$. In this case, the local oscillator problem is analytically solved by taking $f_i=g_{0i}$, and then we get $\eta=r_0$. Generalizing this approach to higher-order $N-$photon states is not straightforward, since there are no general Schmidt decompositions for higher-order tensors. 

The global optimum of our problem therefore corresponds to the eigenpair with the maximum eigenvalue, reducing the problem to finding the leading eigenpair of the tensor $\bm{\psi}$. This is equivalent to computing the \textit{spectral norm} of $\bm{\psi}$\;\cite{friedland2020spectral}, and it is also closely related to the problem of finding its best rank-1 approximation\;\cite{friedland2013best,kofidis2002best,de2000best}, that is,
\begin{equation}
\label{eq:best_rank_1}
\begin{aligned}
\min_{\lambda,\bm{f}} \;  \lVert\bm{\psi}-\lambda \bm{f}^{\otimes N} \rVert_2\quad \text{subject to} \quad \bm{f}^* \bm{f} = 1,
\end{aligned}
\end{equation}
such that $\lambda=\max_{\bm{f}} |\braket{\bm{f}^{\otimes N},\bm{\psi}}|$. This problem has a natural physical interpretation in quantum information theory: the best rank-1 approximation of $\bm{\psi}$ corresponds to its closest separable state, directly connecting our formulation to the geometric measure of entanglement\;\cite{wei2003geometric, ni2014geometric,zhang2020iterative}.

\subsection{\label{sec:eigen_compute}Computation of eigenpairs of symmetric tensors}

The problem of finding eigenpairs of a tensor is nonconvex, and, even in the totally symmetric case, NP-hard beyond order two\;\cite{hillar2013most}. 
The primary source of this difficult can be understood as a consequence of the fact that a symmetric $m-$order tensor $\bm{A}$ can be represented as a degree$-m$ homogeneous polynomial $g(\bm{x})=\bm{A}\cdot\bm{x}^m$\;\cite{brachat2010symmetric}; the unitary eigenvalues correspond to the critical values of $g(\bm{x})$ subject to the normalization constraint (see Eq.\;\eqref{eq:eigenvalue})\;\cite{qi2005eigenvalues,kolda2011shifted}. 
As such, finding eigenpairs is ultimately equivalent to polynomial optimization on a sphere, which is generally nonconvex for degree $m\geq 3$\;\cite{hillar2013most, de2008complexity}. 
Consequently, the optimization landscape typically contains multiple local optima and saddle points. Despite this, practical methods  for computing tensor eigenpairs have been developed.

The most widely used class of methods for computing tensor eigenpairs consists of iterative power-methods, which generalize the matrix power method to higher-order tensors\;\cite{kolda2011shifted,muller2022robust,che2017iterative, de2000best,kofidis2002best}. 
Their update rule generally takes the form
\begin{align}
    \bm{x}_k \mapsto \bm{x}_{k+1}=\frac{\bm{A}\cdot\bm{x}_{k}^{\otimes(m-1)}}{\lVert \bm{A}\cdot\bm{x}_{k}^{\otimes(m-1)}\rVert_2},
\end{align}
where each iteration contracts the tensor with the current iterate (see Eq.\;\eqref{eq:tensor_contraction}) and renormalizes the result. These methods are simple to implement and often effective in practice, but they remain sensitive to initialization and convergence to the global optima is not guaranteed\;\cite{muller2022robust}. To mitigate these limitations,\;\citeauthor{kolda2011shifted} introduced a variant where a shift parameter is incorporated to ensure convergence in a local region\;\cite{kolda2011shifted}. Specialized algorithms of this type have also been developed for complex tensors\;\cite{che2017iterative, zhang2020iterative}.

Beyond power-type iterations, several alternative approaches have been developed, including semidefinite programming\;\cite{cui2014all,nie2014semidefinite,hua2017computing}, homotopy continuation\;\cite{chen2016computing}, and Newton-based methods\;\cite{jaffe2018newton,yu2016adaptive,xu2023feasible,bai2020descent}

Gradient-based methods offer a general and flexible alternative, since they can be applied directly to the polynomial objective on the sphere without structural assumptions on the tensor\;\cite{yu2016adaptive,xu2023feasible,chen2016computing}, and are generally compatible with automatic differentiation frameworks, enabling GPU acceleration\;\cite{jax2018github}. Constraint gradient methods on the sphere generally allow highly accurate results\;\cite{yu2016adaptive,jaffe2018newton,xu2023feasible}, and while they may get trapped in local optima on a nonconvex landscape, they benefit from better theoretical grounding for non-convergence diagnosis compared to iterative power methods\;\cite{kolda2011shifted, bai2020descent,jaffe2018newton}. 

The general dependence on initialization follows from considering $\bm{t} = \sigma_{a}\bm{u}\otimes\ldots\otimes\bm{u} + \sigma_{b}\bm{v}\otimes\ldots\otimes\bm{v}$, an order $l$ totally symmetric tensor,  with $\bm{u}$ orthogonal to $\bm{v}$ and $\sigma_{a} > \sigma_{b}$, and the functional $\bm{t} \cdot\bm{x}^{\otimes l}$ (See Eq. \eqref{eq:eta}), with $x_k=v+\delta_k d_k$. 
At $\bm{v}\otimes\ldots\otimes\bm{v}$, the change in functional value for a small vector $\left(\delta_{1}\bm{d}_{1},\delta_{2}\bm{d}_{2},\ldots,\delta_{l}\bm{d}_{l}\right)$, subject to the condition that $\bm{d}_{l}$ is orthogonal to $\bm{v}$ is
\begin{equation}\label{eq:delta_t}
    \Delta_{\bm{t}} = \sigma_{a}\prod\delta_{k}\left(\bm{u}\cdot\bm{d}_{k}\right)-\sigma_{b}\left(1-\prod_{l}\sqrt{1-\delta_{k}^{2}}\right).
\end{equation}
From this expression, the most positive variation is given by setting $\left(\forall k\right)\;\bm{d}_{k} = \bm{u}$ and $\delta_{k} = \delta$, in which case the relation simplifies to 
\begin{equation}\label{eq:delta_t2}
\Delta_{\bm{t}}  = \sigma_{a}\delta^{l}-\sigma_{b}\left(1-\left(1-\delta^{2}\right)^{l/2}\right).
\end{equation}
When $l = 2$, $\Delta_{\bm{t}} = (\sigma_{a} - \sigma_{b})\delta^2$ indicating gradient search will correctly lead to $\sigma_{a}$, and that gradient approaches can be used to determine the largest singular values in general.  However, for any larger value of $l$, the negative contribution strongly dominates for small $\delta$. 
E.g. if $l = 3$, $\sigma_{a}$ must be approximately $15$ times larger than $\sigma_{b}$ for a step size of $0.1$ (relative to unit vectors) to result in a positive function variation. This highlights that initialization is a fundamental bottleneck across methods: even the added flexibility of gradient-based optimization is not sufficient to overcome a poor starting point, emphasizing the need for a principled initialization strategy, which we develop in the following section.

\section{\label{sec:SVD-informed} SVD-informed method for computing optimal local oscillators }

In this section, we develop an SVD-informed method for computing the optimal local oscillator associated with a photon number state, described by a totally symmetric high-order JSA. We begin by introducing matrix unfoldings of totally symmetric tensors, recalling the higher-order SVD construction, and establishing its connection to the reduced density matrix of a single particle within the $N-$photon state. Building on this, in Sec. \ref{sec:symmetric_decomp} we develop a recursive, rank-1 symmetric decomposition of the tensor and show that the optimal local oscillator is identified with the first maximizer of this decomposition. In Sec. \ref{sec:algorithm} we propose an algorithm to optimize local oscillators that addresses two challenges associated with this problem: the sensitivity to initialization, for which we justify an SVD-based initialization strategy; and the increasing complexity of the optimization landscape as the order of the JSA grows, which we address to develop a homotopy continuation strategy that takes into account the geometry of the JSA. Finally, we provide a convergence bound on the maximal homodyne visibility for multiphoton number states, based on the SVD of bipartitions of the JSA tensor.

\subsection{Preliminaries}

\subsubsection{\label{sec:hosvd}Matrix unfoldings of the totally symmetric tensor and the higher-order singular value decomposition.}

Consider a totally symmetric $N$-order tensor $\boldsymbol{\psi} \in V^N \equiv \mathbb{C}^{I_1\times I_2 \times ... \times I_N}$, with components $\psi_{\scriptscriptstyle i_1 i_2 ... i_N}$. We can reshape this $N$-index object to an ordinary matrix, so that the well-known tools of linear algebra can be applied directly, in particular, the singular value decomposition (SVD).

We define the mode$-k$ unfolding (also called matricization or flattening) of $\bm{\psi}$ as the matrix 
\begin{equation}\label{eq:flattening}
\psi^{(k)}_{\scriptscriptstyle i_k j} \equiv \psi_{\scriptscriptstyle i_k; (i_1 ...,i_{\scriptscriptstyle k-1},i_{\scriptscriptstyle k+1}... i_N)}\in \mathbb{C}^{I_k\times J}
\end{equation} with $J=\prod_{j\ne k}I_j$. In this representation, we keep the index $i_k$ as the row index, while all remaining indices are merged into a single column index $j$\;\cite{kolda2009tensor}, according to 
\begin{align}
    (i_1 ...,i_{\scriptscriptstyle k-1},i_{\scriptscriptstyle k+1}... i_N) \mapsto j=1+\sum_{n \ne k }^N \left(i_n-1\right) \prod_{\scriptscriptstyle m\ne k}^{\scriptscriptstyle n-1}I_m.
\end{align}
This is equivalent to stacking the \textit{mode-k} vectors of the tensor one after the other\;\cite{kofidis2002best,kolda2009tensor}. Because $\bm{\psi}$ is totally symmetric, permuting its indices leaves its entries unchanged, and, consequently, the matricizations obtained from different \textit{modes} are identical; in other words, any \textit{mode$-k$} unfolding yields the same matrix representation. 

Once the tensor is flattened, an ordinary SVD can be applied directly to the matrix $\psi^{(k)}_{\scriptscriptstyle i_k j}$
\begin{align}
    \psi^{(k)}_{i_k j}=\sum_n \lambda_n\, u_{ n\, i_{k}}\, v_{n\, j},
\end{align}
This is the basic building block of the Tucker Higher-Order SVD (HOSVD) procedure introduced by\;\citeauthor{de2000multilinear}\;\cite{de2000multilinear}: an SVD is computed on the matricization along each mode of the tensor,  and the resulting decomposition takes the form
\begin{align}
\psi_{\scriptscriptstyle i_1 i_2 ... i_N}=\sum_{k_1}...\sum_{k_N} \mathcal{G}_{\scriptscriptstyle i_1 i_2 ... i_N} U^{(1)}_{i_1k_1}U^{(2)}_{i_2k_2}... U^{(N)}_{i_Nk_N}, 
\end{align}
where $\mathbf{\mathcal{G}}$ is known as the core tensor and each factor matrix $\mathbf{U^{(k)}}=(u^{(k)}_1,...,u^{(k)}_{r_k})$ is built by stacking the left singular vectors $u^{(k)}_i$ obtained from the SVD of the mode$-k$ unfolding of the general tensor. 

Although this resembles matrix principal component analysis, there is a key structural difference: tensors generally do not admit a decomposition that is simultaneously orthonormal and optimally low-rank. When we enforce orthonormality, as in the HOSVD case, the optimality of the truncated approximation is sacrificed. \\

\subsubsection{Higher order SVD and its relation with the reduced density matrix } 
The connection between the HOSVD and the reduced density matrix of a single particle within the $N-$photon state becomes evident once the flattening is implemented. Defining the density matrix of the full $N-$photon state as $\rho=\ket{\Psi_N}\bra{\Psi_N}$ (see Eq. \eqref{eq:Psi_1}), the reduced density matrix of a single particle is obtained by tracing out the remaining $N-1$ modes,
\begin{align}\label{eq:reduced_rho}
    \rho_I=\text{Tr}_{i_2...i_N}\left(\ket{\Psi_l}\bra{\Psi_N}\right)=
    \sum_{i_1,i'_1} \rho_{i_1,i'_1} a_{i_1}^\dagger \ket{\text{vac}} \bra{\text{vac}}a_{i'_1},
\end{align}
where the discrete frequency-space reduced density matrix is
\begin{align}
    \rho_{i_1i'_1}&=\sum_{j} \psi^{(1)}_{\scriptscriptstyle i_1 j} \,\psi^{(1) *}_{\scriptscriptstyle i'_1 j}.
\end{align}
Substituting the SVD of the flattened JSA tensor, 
\begin{align}
    \psi^{(1)}_{i_1 j}=\sum_l \lambda_l\, u_{\scriptscriptstyle l i_{ 1}}\, v_{\scriptscriptstyle l j},
\end{align}
and using the orthonormality of the right singular vectors,
\begin{align}
\sum_{j}v_{\scriptscriptstyle l j} \,v^*_{\scriptscriptstyle l' j}=\delta_{ll'},
\end{align}
the reduced density matrix takes the diagonal form
\begin{align}
    \rho_{i_1i'_1}=\sum_l \lambda^2_{ l} \,u_{\scriptscriptstyle l i_1} u^*_{\scriptscriptstyle l i'_1},
\end{align}
which resembles a Schmidt decomposition. 
This result shows that the eigenvalues of $\rho_{i_1i'_1}$ are precisely the squared singular values $\{\lambda_l^2\}$ of the mode$-1$ unfolding of the JSA, and the eigenvectors are the corresponding left singular vectors $\{u_{\scriptscriptstyle l i_1}\}$. Moreover, since $\bm{\psi}$ is totally symmetric, the singular values, and hence the eigenvalues of the corresponding single-particle reduced density matrix, are the same for every mode$-k$ unfolding.  

\subsubsection{\label{sec:jsa_rank}The multilinear tensor rank and the optimal local oscillator problem}

The tuple $(r_1,...,r_N)$, where $r_k=\text{rank}(\mathbf{U^{(k)}})$, defines the multilinear rank of the tensor $\bm{\psi}$\;\cite{de2000multilinear,kolda2009tensor} (as defined in Section \ref{sec:hosvd}). If the factor matrix $\mathbf{U^{(k)}}$ computed from the mode$-k$ unfolding is rank-1, then each matrix $\mathbf{U^{(k)}}$ reduces to a vector $\mathbf{u}~\in~\mathbb{C}^{I_k}$, and the Tucker decomposition for a totally symmetric tensor $\bm{\psi}$ reduces to
\begin{align}
\bm{\psi}= \tilde{g} \, \mathbf{u}\otimes \ldots\otimes \mathbf{u}. 
\end{align}
In this case, the core tensor $\mathbf{\mathcal{G}}$ becomes a scalar $\tilde{g}$ and the Tucker decomposition collapses into a rank-1 decomposition. Consistent with the symmetry of the original tensor, this reduced rank-1 approximation is itself totally symmetric and corresponds to a near-best rank-1 symmetric approximation of the tensor $\bm{\psi}$\;\cite{de2000best}. As established earlier, this approximation is also closely connected to the problem of determining the leading eigenpair of the tensor. In this situation, the leading singular vector $\mathbf{u}$ obtained from the HOSVD (or equivalently obtained from the Schmidt decomposition of the single-particle reduced density matrix (Eq. \ref{eq:reduced_rho})) directly solves our optimal local oscillator problem. However, this correspondence does not generally hold for higher-rank tensors. 

\subsection{\label{sec:symmetric_decomp}Symmetric canonical decomposition and the optimal local oscillator}

\noindent
Take $\bm{z_1}^{\otimes {N}} = \bm{z}_{1}\otimes\ldots \otimes\bm{z}_{1}$ to be an element of the unit sphere $\mathbb{S}^{N} = \left\{\bm{f}^{\otimes {N}}\in V^{N}\;\land\;\lVert\bm{f}\rVert_{2}^{2} = 1\right\}$ that maximizes ${\bm{\psi}\cdot\bm{f}^{\otimes{N}}}$ in Eq. \eqref{eq:optim_prob} with value $\lambda_{1}$.
Taking $\left\{\bm{e}_{1},\bm{e}_{2},\ldots,\bm{e}_{n}\right\}$ with $\bm{e}_{1} = \bm{z}_{1}$ to be an orthonormal basis for $V$, $\lambda_{1}$ will be the coefficient of $\bm{e}_{1}\otimes\ldots\otimes\bm{e}_{1}$ in the tensor product expansion of $\bm{\psi}$. Since every such expansion coefficient is related to a pure tensor, $\lambda_{1}$ must be real and larger than any other expansion coefficient in the $\left\{\bm{e}_{1},\bm{e}_{2},\ldots,\bm{e}_{n}\right\}$ basis. 
Further, there is no other basis in which a larger expansion coefficient could be realized. 
\\ \\
Subtracting $\lambda_{1}\;\bm{z}_{1}\otimes\ldots\otimes\bm{z}_{1}$ from $\bm{\psi}$ defines another order$-N$ totally symmetric tensor $\bm{\psi}_{\left(2\right)}$, and so the same argument applies recursively: There is a $\bm{z}_{2}^{\otimes {N}} = \bm{z}_{2}\otimes\ldots \otimes\bm{z}_{2}\in\mathbb{S}^{N}$ that maximizes ${\bm{\psi}_{\left(2\right)}\cdot\bm{f}^{\otimes{N}}}$ with value $\lambda_{2}$; taking $\bm{e}_{1} = \bm{z}_{2}$, $\lambda_{2}$ will be the expansion coefficient of $\bm{\psi}_{\left(2\right)}$ with the largest absolute value (which is in fact real and positive).
\\ \\
By this construction, $\lVert\bm{\psi}_{\left(2\right)}\rVert^{2}_{2}\leq\lVert\bm{\psi}\rVert^{2}_{2}\left(1-\frac{\left(n-1\right)!\;N!}{\left(n + N -1\right)!}\right)$, with $n$ the dimension of the base vector space.
At the $k^{th}$ application of this process 
$$
      \lVert\bm{\psi}_{\left(k\right)}\rVert^{2}_{2}\leq\lVert\bm{\psi}\rVert_{2}^{2}\;\left(1 -\frac{\left(n-1\right)!\;N!}{\left(n + N -1\right)!}\right)^{k}.
$$
Because this expression converges exponentially for any fixed number of dimensions, to arbitrarily good approximation, we may write a finite sum of the form
\begin{equation}
      \bm{\psi} = \sum_{k}\lambda_{k}\;\bm{z}_{k}\otimes\ldots\otimes\bm{z}_{k},
      \label{eq:symExp}
\end{equation}
with $\bm{z}_{k}$ the vector of the $k^{th}$ solution of the maximization problem and  $i\geq j\Rightarrow \sigma_{i}\geq\sigma_{j}$.
\\ \\

As shown by $\bm{\psi} = \lambda_{a}\bm{u}\otimes\bm{u} + \lambda_{b}\bm{v}\otimes\bm{v}$ with $\lambda_{a} >\lambda_{b}$ and $\bm{u}^{\dagger}\bm{v}\neq 0$, the various $\bm{z}_{k}$ vectors appearing in Eq.\;\eqref{eq:symExp} need not be orthogonal. 
The tension between this fact and necessity of the orthogonality in the maximal directions of any flattening of $\bm{\psi}$ highlights a general difficulty associated with finding an optimal local oscillator: finding the best local oscillator amounts to finding $\bm{z}_{1}$, which is not the same as finding the largest singular vector associated with some particular flattening. 
\\ \\

Although any totally symmetric tensor admits a symmetric decomposition as defined in Eq. \eqref{eq:symExp}, this decomposition does not necessarily reflect the tensor's eigenstructure. The first term is an exception: because $\bm{z_1}$ solves a smooth constrained maximization, it must satisfy the associated stationarity condition $\bm{\psi}\cdot\bm{z_1}^{\otimes (N-1)}=\lambda_1 \bm{z_1}$, so $(\bm{z}_1,\lambda_1)$ is guaranteed to be an eigenpair of $\bm{\psi}$, indeed its largest one. This argument does not extend beyond $k=1$: each subsequent $\bm{z}_k$ is obtained by applying the same maximization to the deflated tensor $\bm{\psi}_{(k)}$, and so is only guaranteed to be an eigenvector of $\bm{\psi}_{(k)}$, not of the original tensor $\bm{\psi}$. In fact, for $k\geq2$, tensor eigenvectors of $\bm{\psi}$ do not necessarily appear as elements of a symmetric decomposition, and vice versa. Only in the special case where an eigenvector also appears as one of the generating vectors of a symmetric decomposition the two notions coincide and such eigenvector is known as a robust eigenvector\;\cite{muller2022robust}. \\ \\

\subsection{\label{sec:algorithm}Initialization strategy, homotopy continuation algorithm and bounds}

In the problem of finding the optimal local oscillator of a multi-photon number state, there are two fundamental aspects that complicate the optimization. The first is initialization: as we mentioned in section \ref{sec:eigen_compute}, the performance of general optimization methods for solving the eigenpairs of a tensor depends heavily on their initialization, which is why it is necessary to develop a rigorous initialization strategy. The second is a mismatch of norms: the constraint set for the problem is the unit sphere in the 2-norm, while the objective is order$-N$,  which leads to an ill-conditioned optimization landscape for $N>2$. However, we show that an equivalent formulation using a higher-order norm results in a better conditioned landscape. To address both aspects, we propose an algorithm that initializes the optimization using the leading HOSVD singular vectors and gradually mitigates the norm mismatch using a homotopy continuation strategy. 

To justify this choice of initialization, we return to the basic flattening of the totally symmetric tensor $\bm{\psi}\in V^N$, 
\begin{equation}
      \bm{\psi}^{(k)} = \sum_{k}\sigma_{k}\;\bm{u}_{k}\otimes\bm{v}_{k}.
      \label{eq:basFlt}
\end{equation} 
If $\sigma_{1}\gg\sigma_{2}$, then comparison between Eq.\;\eqref{eq:basFlt} and Eq.\;\eqref{eq:symExp} strongly suggests that $\lambda_{1}\approx\sigma_{1}$ and $\bm{z}_{1}\approx\bm{u}_{1}$.  Directly, $\bm{z}_{1}$ defines the maximal evaluation direction of $\bm{\psi}^{ (k)} $ and
\begin{align}
      \lambda_{1} &= \sum_{k}\sigma_{k} \left(\bm{u}_{k}\cdot\bm{z}_{1}\right)\left(\bm{v}_{k}\cdot \bm{z}_{1}\otimes\ldots\otimes\bm{z}_{1}\right)
      \leq\sigma_{1}.
      \label{eq:valBndBas}
\end{align}
Here each $\bm{v}_{k}$ should be understood as the vectorization of an order$-(N-1)$ symmetric tensor. In Eq. \eqref{eq:valBndBas}, we also used the fact that $\bm{z}_{1}$ is a unit vector, $\lambda_{1}\geq\lambda_{2}\geq\ldots$, and that $\left(\forall k\right)\;\left(\bm{v}_{k} \cdot\bm{z}_{1}\otimes\ldots\otimes\bm{z}_{1}\right)\leq 1$. The same logic applies to the largest singular values of any matrix flattening (i.e. to an arbitrary division of $V^{N}$ into a product of two vector spaces): at best some product of $\bm{z}_{1}$ could perfectly match the largest eigenvector. 

The bound defined previously can be tightened by truncating the sum in Eq. \eqref{eq:valBndBas} to its first $n$ terms. Taking $p_k=\bm{v}_{k}\cdot\bm{z}_{1}\otimes\ldots\otimes\bm{z}_{1}$, $q_k=\bm{u}^{k}\cdot\bm{z}_{1}$, we can define the truncation error
\begin{equation}
    r_n=\sigma_{n+1}\left(\left(1-\sum_{k=1}^n|p_k|^2\right)\left(1-\sum_{k=1}^n|q_k|^2\right)\right)^{1/2}.
\end{equation}
Then, the improvement on Eq. \eqref{eq:valBndBas} is given by 
\begin{equation}\label{eq:bound}
\sum_{k=1}^n \sigma_k \Re\{p_kq_k\}\leq \lambda_1\leq r_n + \sum_{k=1}^n \sigma_k \Re\{p_kq_k\}.    
\end{equation}
Because the accuracy of $r_n$ is set by $\sigma_{n+1}$, Eq. \eqref{eq:bound} quantifies how many of the leading singular vectors are worth including in the initialization. This observation motivates the following initialization strategy. We first form the $\bm{\psi}^{(k)}$ flattening, and compute the first $n$  singular values and left singular vectors. For large tensors this may be accomplished, e.g., by parallelized randomized singular value decomposition. We then run gradient based optimization initialized over the first few left singular vectors--the vectors for which $\sigma_{k}\approx\sigma_{1}$--as well as over linear combinations, recording the optimized value $s_{k}$ and solution $\bm{w}_{k}$. If some $s_{k}$ is sufficiently near $\sigma_{1}$, the associated $\bm{w}_{k}$ is an approximately optimal local oscillator.

As noted above, a substantial portion of the difficulty of the optimization comes from the mismatch of norms. To see this, suppose that the $\bm{v}_{k}$ tensors appearing in the basic flattening were simply outer products of $\bm{u}_{k}$, $\bm{v}_{k}=\bm{u}_{k}^{\otimes N-1}$. Working in the $\left\{\bm{u}_{1},\ldots,\bm{u}_{n}\right\}$ basis would lead to the evaluation formula
\begin{align}
      \bm{\psi}\cdot\bm{f}^{\otimes{N}} = \sum_{k=1}^{n}\sigma_{k} c_{k}^{\ast N},
\end{align}
taking $\bm{f} = c_{1}\bm{u}_{1} + \ldots + c_{n}\bm{u}_{n}$ and $\bm{f}^{\otimes{N}}$ to be the completely symmetric tensor generated by $\bm{f}$. While this problem is not convex under the $2$-norm $\sum_{k=1}^{n} \left|c_{k}\right|^{2} = 1$ constraint, it is convex under the $2N$-norm constraint $\sum_{k=1}^{n} \left|c_{k}\right|^{2N} \leq 1$---simply replacing $c_{k}^{N}$ by $d_{k}$, this is the problem of optimizing against a linear functional over the unit ball. This is the sense in which the diagonal problem is simplified by matching the norm to the JSA order. In general, because $\bm{v}_{k}$ will generally not be a pure tensor, a complete evaluation formula for $\bm{\psi}$ in terms of the $\left\{\bm{u}_{1},\ldots,\bm{u}_{n}\right\}$ basis contains a large number of additional cross terms that exhibit scaling ranging from linear to $N$. However, because every term is homogeneous, the optimization remains convex under the $2N$-norm under the non-linear transformation $c_{k}^{N} = d_{k}$. We take advantage of this fact by developing an optimization algorithm in which the unit-norm constraint is enforced through an exterior penalty combined with a homotopy continuation\;\cite{chen2016computing, kuo2018continuation} on the constraint norm. Here, we convert the constrained problem into the unconstrained objective
\begin{align} \label{eq:penalty_continuation}
    \max_{\bm{f}} \left(\bm{\psi} \cdot \bm{f}^{\otimes N} - \alpha_k \left(\|\bm{f}\|_{p_k} -1\right)^2\right),
\end{align}
where $\bm{\psi} \cdot \bm{f}^{\otimes N}$ denotes the homodyne overlap as defined in Eq. \eqref{eq:discrete_eta},  $\alpha_k$ is the penalty weight, and $p_k$ is the norm exponent at outer iteration $k$. The norm exponent is ramped from an initial value $p_0=2N$ down to the target value $p=2$ according to a linear schedule over $n_{\mathrm{stages}}$,\begin{align}
    p_k=2N+\frac{k}{n_{\mathrm{stages}}-1}(2-2N), \quad k=0,...,n_{\mathrm{stages}}-1,
\end{align}
where $N$ is the order of the JSA tensor. Starting from the smoother $2N-$norm, whose penalty landscape is better conditioned, and progressively approaching the $2-$norm allows the optimizer to track a continuous family of solutions instead of dealing with the nonconvexity of the $2-$norm from the beginning. The solution at each stage $k$ is used as the warm start for the next, implementing a continuation strategy over the constraint.

\subsubsection{\label{sec:knapsack_bounds}  Bounds}

Take $\bm{\psi}$ to be a completely symmetric JSA tensor of order $N$. In Section \ref{sec:hosvd} we introduce the mode$-k$ flattening $\bm{\psi}^{(k)}$, obtained by grouping a single index $i_k$ against the remaining $N-1$ indices. More generally, one can split the $N$ indices of $\bm{\psi}$ into two disjoint sets in many ways, for example, by grouping any $l$ indices against the remaining $m=N-l$ indices. The SVD of the resulting matricization takes the form $\bm{\psi}^{(l,m)}=\sum_k \sigma_k u_k^{[l]}\otimes v_k^{[m]}$, where $u_k^{[l]}$ and $v_k^{[m]}$ are the left and right singular tensors of order $l$ and $m$ associated with this bipartion. Using this more general bipartition of $\bm{\psi}$, the optimization problem defined in Eq. \eqref{eq:optim_prob} can be rewritten as 
\begin{align}
&\max_{\bm{f}}\;\sum_{k}\sigma_{k}\;\left(\bm{u}_{k}^{[l]}\cdot \bm{f}^{\otimes l}\right)\left(\bm{v}_{k}^{[m]}\cdot \bm{f}^{\otimes m}\right) \label{eq:mainOptFlat} \\
&\;\text{such that}~\lVert\bm{f}\rVert_{2} = 1,
\nonumber
\end{align}
with $N = l + m$. From the starting point of Eq.\;\eqref{eq:mainOptFlat} a variety of bounds can be derived, and determining the tightest result invariably requires exploring a number of branching options.

By defining the coefficients
\begin{equation}\label{eq:r_k}
    r_k=\left|\bm{u}_{k}^{[l]}\cdot \bm{f}^{\otimes l}\right|, \quad
    s_k=\left|\bm{v}_{k}^{[m]}\cdot \bm{f}^{\otimes m}\right|,
\end{equation}
Eq. \eqref{eq:mainOptFlat} is bounded by
\begin{align} \label{eq:boundExpression}
&\max_{\bm{s},\,\bm{r}}\;\sum_{k}\sigma_{k}r_{k}s_{k}~~\text{such that}\\
&\;~\forall k\;\left(0\leq r_{k}\leq\mu_{k}\right),\;\&\;
\sum_{k} r_{k}^{2} \leq 1,
\nonumber\\
&\;~\forall k\;\left(0\leq s_{k}\leq\nu_k\right),~
\sum_{k} s_{k}^{2} \leq 1,
\nonumber 
\end{align}
where $\mu_k$ and $\nu_k$ are defined as
\begin{align}
	\mu_{k} &= \max_{\bm{f}}\;\left|\bm{u}_{k}^{[l]}\cdot \bm{f}^{\otimes l}\right|\;\text{such that}~\lVert\bm{f}\rVert_{2} = 1,
	\label{eq:munuDef} \\
	\nu_{k} &= \max_{\bm{f}}\;\left|\bm{v}_{k}^{[m]}\cdot \bm{f}^{\otimes m}\right|\;\text{such that}~\lVert\bm{f}\rVert_{2} = 1.
	\nonumber
\end{align}

A particular case worth mentioning is the case of $n$ even. In this case, consider  $l=m=n/2$, then, by the total symmetry of the JSA tensor $\bm{\psi}$, Eq. \eqref{eq:mainOptFlat} becomes
\begin{align}
&\max_{\bm{f}}\;\sum_{k}\sigma_{k}\;\left(\bm{u}_{k}^{[n/2]}\cdot \bm{f}^{\otimes n/2}\right)\left(\bm{v}_{k}^{[n/2]}\cdot \bm{f}^{\otimes n/2}\right) \label{eq:mainOptEven} \\
&\;\text{such that}~\lVert\bm{f}\rVert_{2} = 1,
\nonumber
\end{align}
and Eq. \eqref{eq:boundExpression} becomes 
\begin{align}
	&\max_{\bm{s},\,\bm{r}}\;\sum_{k}\sigma_{k}s_{k}^{2}~~\text{such that} 
	\label{eq:boundExpressionEven} \\
	&\;~\forall k\;\left(0\leq s_{k}\leq\nu_{k}\right),~\sum_{k} s_{k}^{2} \leq 1.
	\nonumber
\end{align}

When $l = 2$ (resp. $m = 2$), by Takagi-Autonne (eigenvalue) decomposition, $\mu_{k}$ (resp. $\nu_{k}$) is bounded by $\sigma_{1}\left(\bm{u}_{k}^{[2]}\right)$ (resp. $\sigma_{1}\left(\bm{v}_{k}^{[2]}\right)$), with $\sigma_{1}\left(\ldots\right)$ denoting the largest singular value of the enclosed order two tensor. 
When $l = 1$ (resp. $m=1$), $\mu_{k} = 1$ (resp. $\nu_{k} = 1$). In all other cases Eq.\;\eqref{eq:munuDef} is a restatement of either Eq.\;\eqref{eq:mainOptFlat} or Eq.\;\eqref{eq:mainOptEven} on a lower order tensor, and is bounded, on a lower level, by following the same sequence of steps until one of the base cases, a tensor of order $2$ or $1$, is reached.

As a consequence of Young's inequality, the objective for Eq.\;\eqref{eq:boundExpression}, for any given values of $\tau_{k} > 0$, is bounded by
$$
\sum_{k}\sigma_{k}r_{k}s_{k} \leq \sum_{k}\sigma_{k}\left(\tau_{k}r_{k}^{2} + \tau_{k}^{-1}s_{k}^{2}\right)/2.
$$
Inserting this relaxation, Eq.\;\eqref{eq:boundExpression} splits the optimization into a pair of fractional knapsack problems. Similarly for Eq. \eqref{eq:boundExpressionEven}. Optimization belonging to the fractional knapsack class are simply solved by making the coefficient associated with the largest value of $\sigma_{k}\tau_{k}$ as large as possible subject to the component bound,  and then moving to the next best option until the sum bound is reached.
To determine a sensible bound, a minimization must be done over the convex problem posed by the $\tau_{k}$ variables.
Because a maximum over a collection of convex functions is convex, this optimization is guaranteed to have a unique minimum, which can be found by bisection, fixed point iteration (or essentially any other optimization algorithm).
\\ \\

\section{\label{sec:results} Numerical experiments}

\subsection{Setup}

\subsubsection{\label{sec:jsa_simulation} Simulation of the joint spectral amplitudes of photon number states with controlled correlation structure}

To evaluate the proposed method, we consider two families of JSAs that arise naturally in the context of Fock state generation in quantum photonics: multivariate Gaussian JSAs and the JSAs associated with Fock state generation via high-order parametric down-conversion. Both families allow systematic control over their spectral correlation structure by varying some experimental parameters that we define below, which make it possible to rigorously study how the optimal local oscillator depends on the spectral entanglement of the respective photon number state. 

\paragraph{\label{sec:gaussian_jsa}Gaussian JSAs:} Gaussian JSAs arise in waveguided sources engineered to produce spectrally pure photon pairs\;\cite{quesada2018gaussian,poveda2023} and their higher-order generalizations. Some fabrication strategies commonly used to generate Gaussian JSAs include periodic poling of the nonlinear medium\;\cite{graffitti2018design}, dispersion engineering\;\cite{uren2005pure,fang2013state,chen2017efficient,osorio2025strategies,malik2026high}, and filtering.

We model the $N-$photon JSA as a multivariate normal density function,
\begin{equation}\label{eq:multivar_normal}
\phi(\bm{\omega}) = \phi_0
\exp\left( -\frac{1}{2} (\bm{\omega} - \boldsymbol{\omega_0})^\top 
\boldsymbol{\Sigma}^{-1} (\bm{\omega} - \boldsymbol{\omega_0}) \right),
\end{equation}
where $\phi_0$ is a normalization constant, $\boldsymbol{\omega_0} \in \mathbb{R}^N$ is the mean frequency vector, $\Sigma \in \mathbb{R}^{N\times N}$ is the covariance matrix, and $\bm{\omega}=(\omega_1,\omega_2,...,\omega_N)$. The covariance matrix is positive definite and symmetric; its diagonal entries $\sigma_{ \omega_i,\omega_i}$ represent the variances of each frequency variable,  and the off-diagonal entries $\sigma_{\omega_i,\omega_j} $ represent the pairwise spectral covariances. The correlation matrix $\mathbf{R}$ is related to $\mathbf{\Sigma}$ through the transformation $\mathbf{\Sigma}=\mathbf{DRD}$, where $\mathbf{D}=\sqrt{\text{diag}(\mathbf{\Sigma})}$. The correlation matrix is equivalent to the covariance matrix of standardized random variables $r_{\omega_i,\omega_j}=\sigma_{\omega_i,\omega_j}/\sigma_{\omega_i}\sigma_{\omega_j}$, where $\sigma_{\omega_i}=\sqrt{\sigma_{\omega_i,\omega_i}}$ is the standard deviation of $\omega_i$. 

Considering the total symmetry of photon-number JSAs, we impose permutational symmetry among the $N$ frequency variables, so that all off-diagonal entries of $\mathbf{\Sigma}$ are equal to a common Pearson coefficient $r$, so that $r_{\omega_i,\omega_j}=r_{\omega_j,\omega_i}=r$. In this case, the covariance matrix in standardized variables has the form
\begin{equation}
\Sigma = \begin{pmatrix}
1      & r      & \cdots & r \\
r      & 1      & \cdots & r \\
\vdots & \vdots & \ddots & \vdots \\
r      & r      & \cdots & 1
\end{pmatrix}
\end{equation} and its eigenvalues are $\lambda_1=1+(N-1)r$ and $\lambda_2=\lambda_3=...=\lambda_N=1-r$. Positive definiteness and invertibility of $\mathbf{\Sigma}$ require $-\frac{1}{N-1}<r<1$. 

\begin{figure}[h]
    \includegraphics[width=0.5\textwidth]{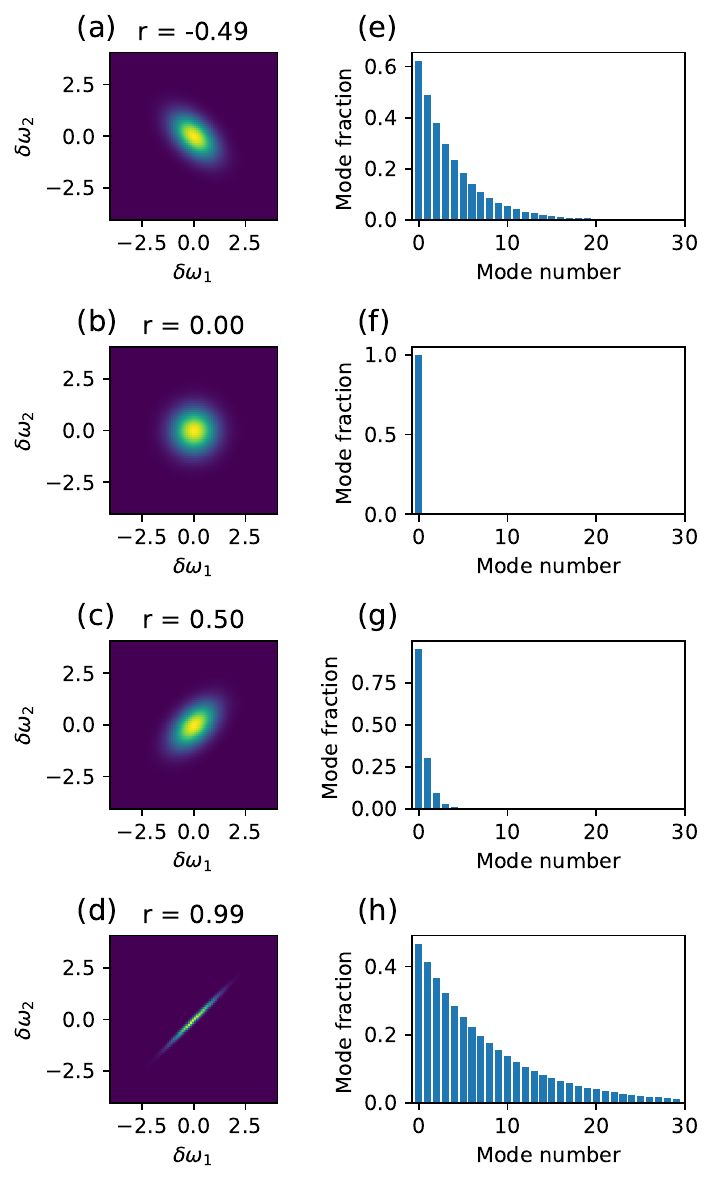}   \caption{\label{fig:proj_mode_decomp} (a-d) Intensity projections of three-dimensional Gaussian JSAs in the plane $\{\delta\omega_1,\delta\omega_2\}$ with varying correlation degrees ($r:$ Correlation parameter). (e-h) Mode decomposition obtained from the HOSVD, each paired with the JSA directly to its left.}
\end{figure}

The parameter $r$ directly controls the degree of spectral entanglement among the emitted photons. In the limit $r \to -\frac{1}{N-1} ^{+}$, the system is maximally anti-correlated, and as $r\to 1^{-}$, the system approaches the perfectly correlated case. The case where $r=0$ corresponds to spectrally uncorrelated photons and yields a separable JSA. To illustrate this behavior, Fig. \ref{fig:proj_mode_decomp}(a-d) shows intensity projections of three-dimensional Gaussian photon-triplet JSAs $\phi(\omega_1,\omega_2,\omega_3)$ with varying correlation parameters, displayed in the plane $\{\omega_1,\omega_2\}$ and centered around the central frequencies $\delta\omega_i=\omega_i-\omega_{i0}$. In Fig.\ref{fig:proj_mode_decomp}(e-h) we plotted the spectral mode distributions obtained from the HOSVD procedure. 
From the perspective of the optimization problem, varying $r$ tunes the dominance of the leading eigenpair of the JSA tensor: weakly correlated states ($r\approx0$) produce a well-defined global optimum, whereas anti-correlated and highly correlated states increase the difficulty of the optimization. 

\paragraph{\label{sec:hopdc_jsa}JSAs associated with the generation of photon-number states by high-order parametric down-conversion: } A distinct class of JSAs arises in high-order parametric down-conversion (HOPDC), a nonlinear optical process in which a single pump photon is annihilated to produce $N$ daughter photons. This process is the generalization of spontaneous parametric down-conversion, a second order nonlinear process commonly used to generate heralded single photons and photon pairs. This process has been studied theoretically as a source of multipartite entangled states\;\cite{okoth2019seeded, banic2022resonant,osorio2025strategies}. Unlike the Gaussian case, these JSAs have a richer geometrical structure, presenting a more challenging test for spectral optimization.

The JSA for $N-$photon SPDC in a waveguide of length $\ell$ takes the form
\begin{align}\label{eq:HOPDC_JSA}
\psi(\bm{\omega})=\psi_0 ~\varphi_{\text{eff}} \left(\bm{\omega}\right) \alpha\left(\textstyle\sum_i^N \omega_i\right),
\end{align}
where $\bm{\omega}=(\omega_1,\omega_2,...,\omega_N)$ is the frequency vector, $\psi_0$ is a normalization constant, $\varphi_{\text{eff}}$ is an effective phase-matching function, and $\alpha$ is the pump envelope function. The effective phase-matching function incorporates both the waveguide's phase-matching response and the spectral transmission of the system.

We assume a Gaussian pump envelope,
\begin{align}\label{eq:pump}
\alpha(\bm{\omega})=\alpha_0 \exp\left(-\frac{1}{2\sigma^2}\left(\sum_i \left(\omega_i-\omega_{i0}\right)\right)^2\right),
\end{align}
where $\alpha_0$ is a normalization constant such that $\int d\bm{\omega}|\alpha(\bm{\omega})|^2=N_P$, with $N_P$ being the average number of pump photons and $\sigma$ is the bandwidth of the pump.

The effective phase-matching function $\varphi_{\text{eff}}$ is defined as
\begin{align}\label{eq:pm_eff}
\varphi_{\text{eff}} =\text{H}(\bm{\omega})~\text{sinc} \left(\frac{\ell}{2} \Delta k(\bm{\omega})\right),
\end{align}
where $\text{H}(\bm{\omega})$  is a filter function. Here, we specialize to the degenerate generation regime, in which all $N$ output photons are produced at the same central frequency $\omega_{F0}$ in the same spatial mode, so that the pump is centred at $\omega_{P0}=N\omega_{F0}$.  Then, if we consider a waveguide that is homogeneous longitudinally, the phase-mismatch is
\begin{align}\label{eq:phase_mismatch}
    \Delta k(\bm{\omega})=k_P\left(\sum_i^N \omega_i\right) -\sum_i^N k_F(\omega_i),
\end{align}
where $k_{P,F}(\omega_i)=\omega_i~n_{P,F}(\omega_i)/c$, with $n_{P,F}(\omega_i)$ being the effective refractive index of the pump ($P$) or $N-$photon signal ($F$) modes at the frequency $\omega_i$, and $c$ being the velocity of light in vacuum. In this case, the filter function has the form ${\text{H}(\bm{\omega})=\textstyle\prod_i^N h(\omega_i)}$, where $h(\omega)$ defines the transmission amplitude of the filter, which may represent material absorption, propagation loss, or an external spectral filter applied at the output of the nonlinear medium.

In terms of the waveguide's dispersion parameters, we expand the wavenumbers in Eq. \eqref{eq:phase_mismatch} in a Taylor series around the central frequencies $\omega_{P0}$ and $\omega_{F0}$, respectively. We keep terms up to second order, assuming we work in the regime where higher-order dispersion terms are negligible. We also neglect other nonlinear effects, such as self- and cross-phase modulation. Defining $\delta\omega_i=\omega_i-\omega_{F0}$,
the phase-mismatch is
\begin{align}\label{eq:pm_taylor}
   \Delta k(\bm{\omega})=\, \left(k_{P0} -N \, k_{F0}\right)+\left(\frac{1}{v_P} -\frac{1}{v_F}\right)\sum_i^N \delta\omega_i \nonumber \\
   + \tfrac{1}{2}\beta_P\left(\sum_i^N \delta\omega_i\right)^2  -\tfrac{1}{2}\beta_F \sum_i^N \delta\omega_i^2.
\end{align}
The parameters $k_{P0,F0}$, $v_{P,F}$, and $\beta_{P,F}$ correspond to the wavenumber, group-velocity, and group-velocity dispersion, each evaluated at the central frequencies of the pump and $N-$photon signal fields.  

\begin{figure}[h]
\includegraphics[width=0.45\textwidth]{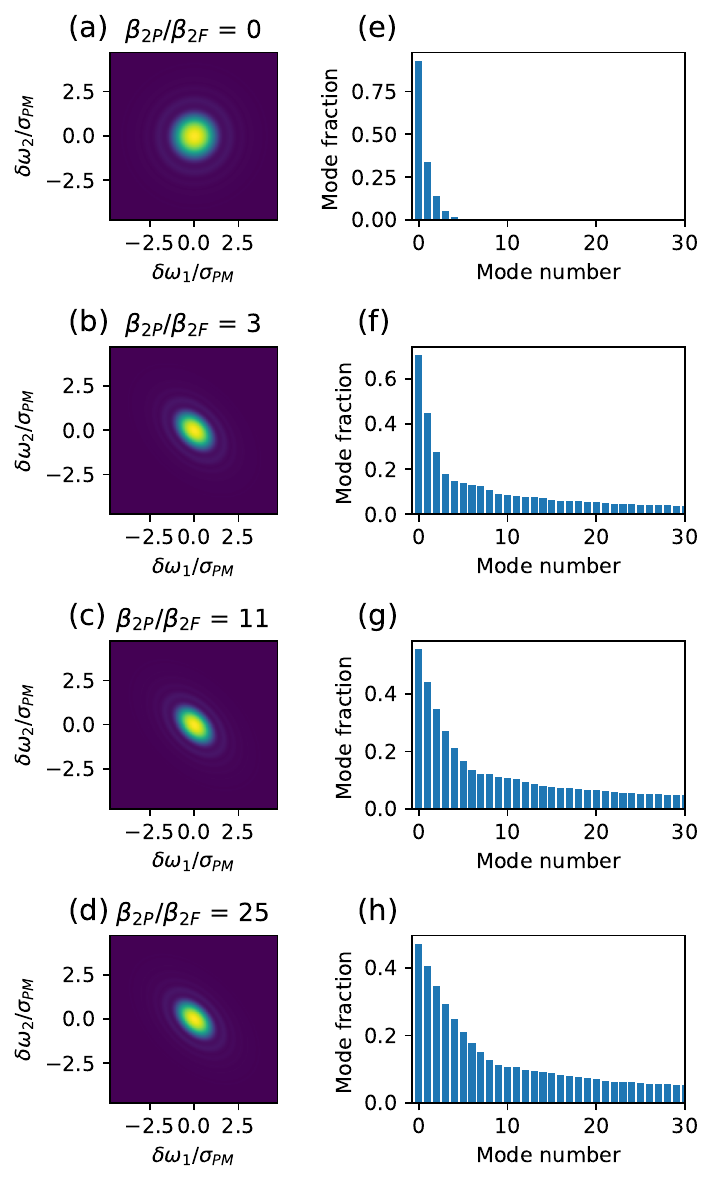}
\caption{\label{fig:proj_hopdc_mode_decomp} (a-d) Intensity projections of JSAs associated with photon-number states generated through HOPDC varying the ratio between the pump's and $N-$photon signal's group-velocity dispersion parameters $\beta_{2P}/\beta_{2F}$. The frequency variables are normalized to the bandwidth $\sigma_{PM}$ of each JSA. (e-h) Mode decomposition obtained from the HOSVD, each paired with the JSA directly to its left.}
\end{figure}

In Fig. \ref{fig:proj_hopdc_mode_decomp}(a-d), as an example, we show intensity projections of three-dimensional HOPDC JSAs onto the plane $\{\delta\omega_1,\delta\omega_2\}$, varying the ratio between the pump's and $N-$photon signal's group-velocity dispersion parameters, $\beta_P$ and $\beta_F$, respectively. The frequency variables are normalized by the bandwidth of the phase-matching function $\sigma_{PM}\approx\sqrt{4\pi/\ell|\beta_F|}$. We considered an ultrafast pump bandwidth with $\sigma> \sigma_{PM}$ and assumed perfect phase-matching ($k_{P0} -N \, k_{F0}$) and group-velocity matching ($v_P\approx v_F$). 

\subsubsection{Optimization methods}

We consider photon-number states of order $N\in\{3,4,5,6\}$ considering Gaussian JSAs and $N\in\{3,4,5\}$ for JSAs associated to HOPDC. This range is chosen because $N=2$ admits an analytic ground truth through the standard Schmidt decomposition, while computational feasibility constrains the upper limit for the grid-based experiments from our proposal. A direction for future work is to exploit the permutation symmetry of the JSAs to extend these methods to higher photon numbers, decreasing memory usage and computational cost. 

\paragraph{Initialization: } For each JSA tensor, we compute its mode$-1$ matrix flattening by reshaping the tensor using standard array operations. Because the JSA is totally symmetric, all mode$-k$ matricizations are equivalent. Consequently, we only compute one matrix flattening rather than all possible unfoldings. For an $N-$mode symmetric tensor of shape $m \times m \times \dots \times m$, the mode$-1$ flattening has dimensions $m \times m^{N-1}$. 
We then apply randomized SVD\;\cite{tensorrsvd} to the resulting matrix and keep its leading left singular vectors as the initialization points for the optimizer. Randomized SVD is used in place of an exact SVD because the latter becomes computationally expensive as $N$ and the frequency-grid resolution increases. The randomized variant provides a high accuracy computation speedup for the leading singular vectors.

\paragraph{Gradient-based optimization with adaptive penalty and norm homotopy: } All optimizations are implemented in Python using the library JAX\;\cite{jax2018github}, which provides just-in-time compilation and automatic differentiation of the objective and constraint. The core optimizer is L-BFGS-B, a quasi-Newton method. The unit-norm constraint is enforced through an exterior penalty combined with a homotopy continuation\;\cite{chen2016computing} on the constraint norm, as described in Section\;\ref{sec:algorithm}. For the schedule described above in Eq.\;\eqref{eq:penalty_continuation},   the penalty weight $\alpha_k$ starts small (e.g., $10^{-2}$) and is increased across iterations according to a two-phase adaptive schedule. In the initial phase, while the constraint violation $|\|\bm{f}\|_{p_k} -1|$ exceeds a tolerance value $\delta_{\text{viol}}$ (e.g.,$10^{-2}$), the weight is updated by multiplying it by a small factor as $\alpha_{k+1}=\kappa \cdot \alpha_k$, e.g., with $\kappa_{\text{init}}=2$. This keeps the optimization stable while the solution is still far from feasible. Once the violation falls below $\delta_{\text{viol}}$, the schedule switches to a tightening phase with $\kappa_{\text{tight}}=5$. This accelerates convergence to a nearly feasible solution. The optimization finishes when both the constraint violation and the relative change in objective $\Delta=|\eta(\bm{f_{\text{new}}})-\eta(\bm{f})|$ fall below a small tolerance value $\varepsilon_{\text{obj}}$ (e.g.,$10^{-6}$). Since the $2-$normalization of $\bm{f}$ and the JSA bound the objective by 1, the final tolerance  $\varepsilon_{\text{obj}}$ corresponds to  a convergence precision in absolute terms.

\paragraph{Baseline and success criterion: } As a baseline we use Trust-Region Basin Hopping (TRBH), a global optimization scheme that alternates between trust-region local minimization and small perturbations to the current iterate, allowing the algorithm to move from local minima and explore the optimization landscape. However, given its rapidly growing computational cost with JSA dimensionality and grid resolution, TRBH is not practical as a general purpose optimizer for our problem; we therefore apply it only to a representative sample of JSAs from the full set, using it as a verification tool.  Details are provided in the Appendix\;\ref{sec:comp_perform}. 

For the full set of experiments, we apply the following success criteria: first, we keep the solution with the highest overlap $\eta$ across all initializations, and second, we confirm that the solution follows the tensor eigenvalue condition. As an additional consistency check, we compare this optimum against the HOSVD upper bound derived in Section \;\ref{sec:knapsack_bounds}.

\subsection{Optimization results}

\subsubsection{Motivating example}

\begin{figure}[ht]
    \includegraphics[width=0.23\textwidth]{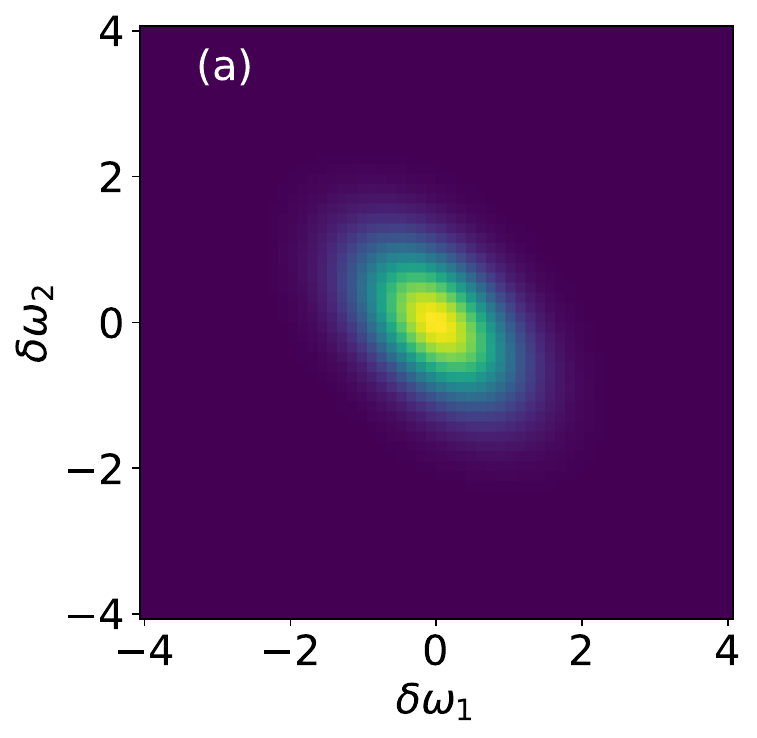}%
    \includegraphics[width=0.23\textwidth]{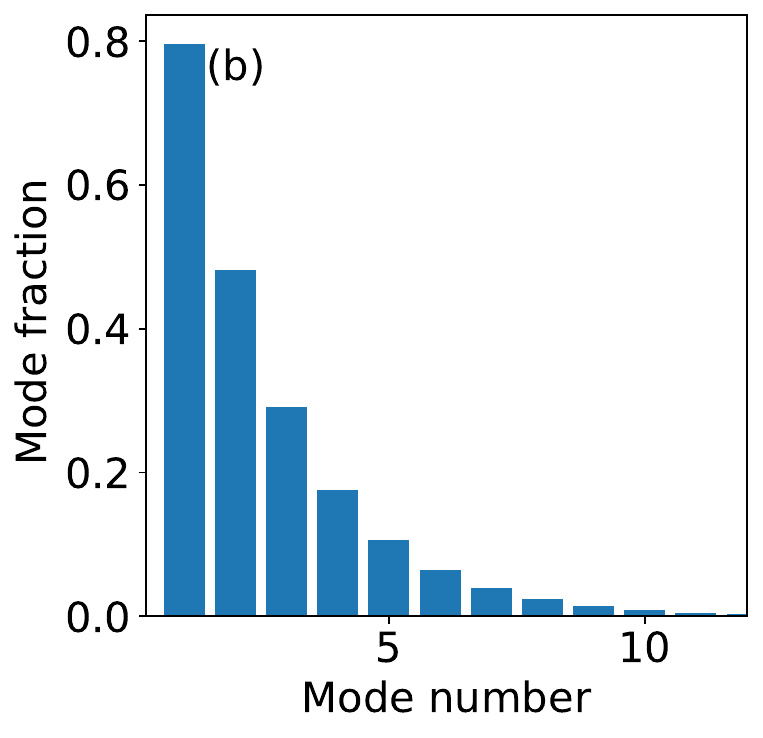}%
    \\
    \includegraphics[width=0.225\textwidth]{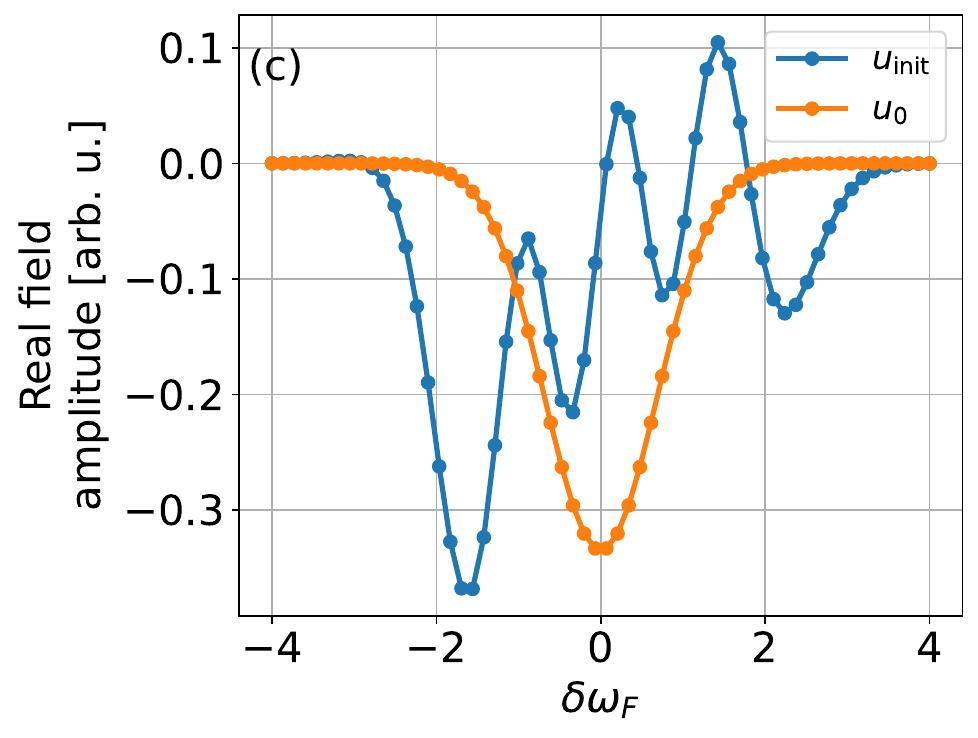}%
    \includegraphics[width=0.235\textwidth]{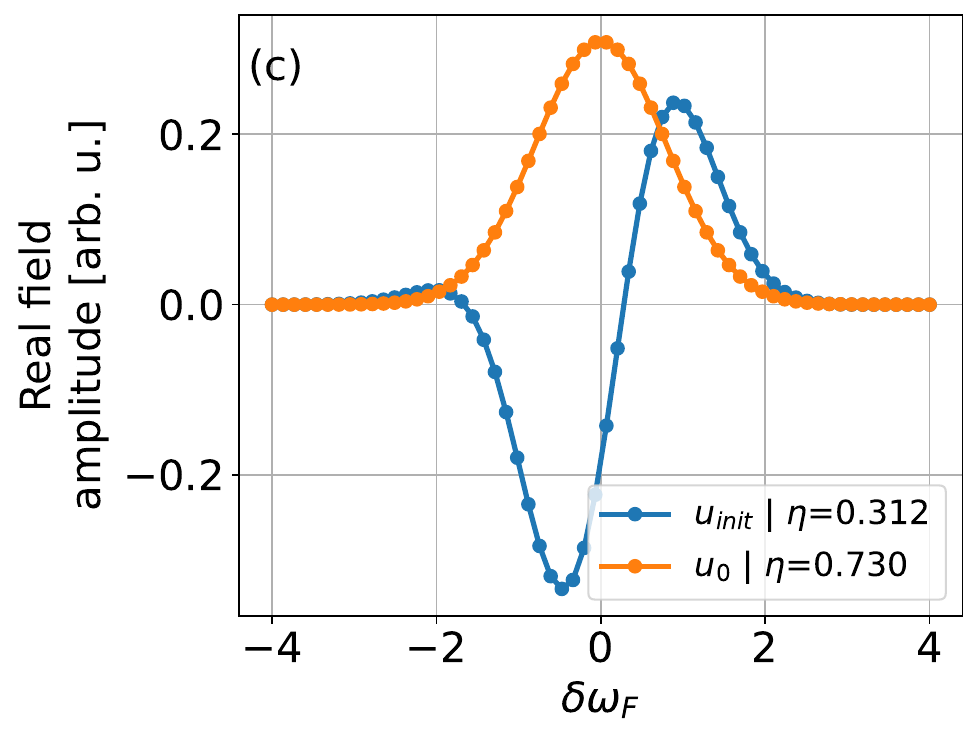}%
    \caption{\label{fig:motiv_example} (a) 2D projection of a Gaussian photon triplet JSA with correlation parameter $r=-0.46$. (b) Mode distribution from the HOSVD of the JSA. (c) Initializing
distributions:  $u_0$: Leading HOSVD vector, $u_{\text{init}}$: Linear combination of 8 first HOSVD vectors. (d) Optimized local oscillators. The routine initialized with $u_{\text{init}}$ converges to a local maximum. }
\end{figure}

Before presenting our general optimization results, we illustrate with a concrete example why the choice of initialization is crucial for this problem. For this, let's consider a three-dimensional Gaussian photon triplet JSA $\phi(\omega_1,\omega_2,\omega_3)$ with correlation parameter $r=-0.46$, projected onto the plane $\{\omega_1,\omega_2\}$ and centered around the central frequencies $\delta\omega_i=\omega_i-\omega_{i0}$ in Fig.;\ref{fig:motiv_example}(a). This value of $r$ is representative of a moderately anticorrelated source, far from the separable limit $r=0$, and, as we will see, it already produces a landscape with distinguishable local maxima.

The HOSVD of the JSA tensor yields an ordered set of candidate LO spectral modes, ranked according to the singular values of the flattened tensor. Fig.\;\ref{fig:motiv_example}(b) shows the distribution of spectral weights among the leading modes: although the first singular vector holds a large proportion of the total mode fraction, the remaining modes still carry non-negligible weight, indicating that the JSA is not well approximated by a rank-1 tensor. This multimode structure is an indication of the non-convexity of the optimization landscape and is a signal of entanglement among the spectral modes of this particular photon triplet state. 

We compare two initializations for the gradient-based optimizer. The first is the leading HOSVD singular vector $\bm{u_0}$. The second corresponds to a linear combination of the first eight higher-order singular vectors $\bm{u_{\text{init}}}= \sum_{i=0}^7 \bm{u_i}/ \lVert \sum_{i=0}^7 \bm{u_i}\rVert_2$. Both initializations are shown in Fig.\ref{fig:motiv_example}(c). Starting from $\bm{u_0}$, the optimizer converges to a local oscillator with nonlinear overlap $\eta_1=0.730$, which we identified as the global maximum based on the search over the HOSVD grid proposed in Section \ref{sec:algorithm} and confirmed against a set of random initializations. Starting from the higher-order combination $\bm{u_{\text{init}}}$, the optimizer converges to a distinct mode profile shown in Fig. \ref{fig:motiv_example}(d) and a smaller  nonlinear overlap $\eta_{\text{init}}=0.312$.

This example makes two points concrete. First, the optimization landscape is genuinely nonconvex: different initializations converge to different stationary points, and gradient descent alone provides no guarantee of finding the global optimum. Second, a robust initialization strategy is needed to certify optimality. It's recommended to evaluate multiple candidate initializations, to then compare and keep the best converged solution. The leading HOSVD singular vectors are natural candidates for the initialization grid because they are easy to compute, and they generally reflect the dominant spectral structure of the state.

\subsubsection{\label{sec:numerical_eigen}Numerical validation of the eigenvalue condition}

The theoretical analysis of Sections\;\ref{sec:quadrature} and\;\ref{sec:optim_problem} establishes that the optimal local oscillator distributions satisfy the eigenvalue condition (Eq. \eqref{eq:eigenvalue}). We now verify this characterization numerically across both JSA families and all tested dimensions.  

\begin{figure}[h]
\includegraphics[width=0.23\textwidth]{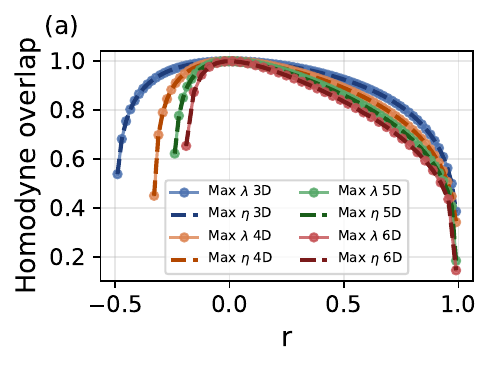}
\includegraphics[width=0.23\textwidth]{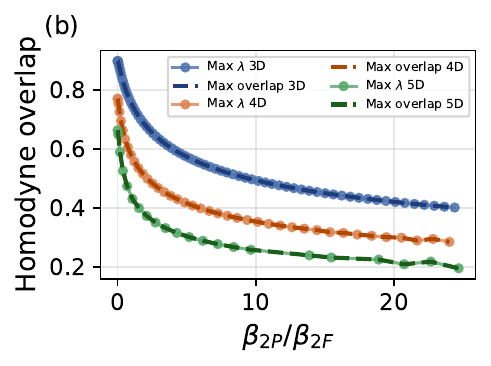}
\caption{\label{fig:max_overlap}
 Maximum homodyne overlap $\eta$ per correlation parameter $r$ for (a) Gaussian JSAs with $N\in\{3,4,5,6\}$ and for different ratios $\beta_{2P}/\beta_{2F}$ for (b) HOPDC JSAs, with $N\in\{3,4,5\}$. The dashed curves represent the maximum eigenvalue $\lambda$ for each JSA.  }
\end{figure}

Figure\;\ref{fig:max_overlap}(a) shows the optimal homodyne overlap $\eta$ for Gaussian JSAs as a function of the correlation parameter $r$, and Figure\;\ref{fig:max_overlap}(b) shows the optimal overlap $\eta$ for JSAs associated to HOPDC in terms of the ratio between the pump's and $N-$photon signal's group-velocity dispersions $\beta_{2P}/\beta_{2F}$, as defined in Section \ref{sec:hopdc_jsa}. Both $r$ and $\beta_{2P}/\beta_{2F}$ affect the multilinear tensor rank of the JSA (See Section\;\ref{sec:jsa_rank}). Alongside, we plotted the leading eigenvalue $\lambda$ for each JSA computed independently. 

Across all dimensions and both the Gaussian and HOPDC families, the two quantities are in agreement, confirming that the optimization converges to the true leading eigenpair. A notable feature of these curves is that the highest overlaps are achieved in the low-correlation regime ($r$ near-zero or $\beta_{2P}\ll\beta_{2F}$), and the homodyne overlap degrades as $r$  and the ratio $\beta_{2P}/\beta_{2F}$ increase in magnitude. As discussed in the introduction, the overlap $\eta$ directly controls the efficiency with which homodyne detection can resolve the non-Gaussian features of the photon number state; the degradation with correlation therefore implies that highly correlated states present a greater challenge for a homodyne characterization of the quantum state. 

\begin{figure}[h]
\includegraphics[width=0.23\textwidth]{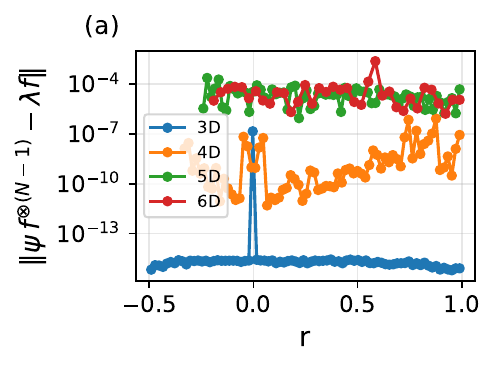}
\includegraphics[width=0.23\textwidth]{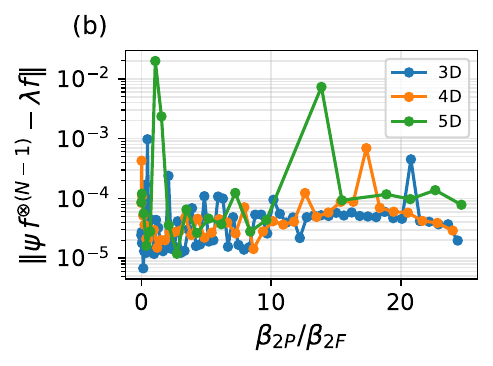} 
\caption{\label{fig:fidelity}  Residual eigenvalue condition $\lVert\bm{\psi} \cdot \bm{f}^{\otimes(N-1)}-\lambda\bm{f}\rVert_2$ considering the optimal LO distributions $\bm{f}$ obtained from our optimization algorithm for (a) Gaussian JSAs, sweeping across the correlation parameter $r$ at $N\in\{3,4,5,6\}$, and (b) HOPDC JSAs, with $N\in\{3,4,5\}$, sweeping across different ratios $\beta_{2P}/\beta_{2F}$. }
\end{figure}

To validate the eigenvector condition directly, Figure\;\ref{fig:fidelity} presents the residual $\lVert\bm{\psi} \cdot \bm{f}^{\otimes(N-1)}-\lambda\bm{f}\rVert_2$ evaluated at the converged solutions. For both JSA families and all dimensions $N$, the residuals remain small throughout the entire valid range of $r$ and $\beta_{2P}/\beta_{2F}$. This confirms that the converged solutions are true eigenpairs of each respective JSA. The achievable residual is governed by the tolerances of the gradient descent algorithm, specifically, the stationarity precision on the gradient of the objective function ($\left|\nabla\eta_\lambda\right|<10^{-8}$) and the normalization constraint ($\left|\bm{f}^\ast \bm{f}-1\right|<10^{-6}$). Consequently, the residuals exhibit a mild growth with tensor order $N$: for $N=3$ in Gaussian JSAs the residuals approach machine precision, while higher orders yield larger values, as the accumulation of floating-point operations in the contraction $\bm{\psi} \cdot \bm{f}^{\otimes(N-1)}$ compounds the effect of the finite solver tolerance. For applications requiring tighter eigenpair accuracy, the residual condition $\lVert\bm{\psi} \cdot \bm{f}^{\otimes(N-1)}-\lambda\bm{f}\rVert_2<\epsilon$ can be added to the other stopping criteria. 

\subsubsection{Bound evaluation}

\begin{figure}[h]
\includegraphics[width=0.23\textwidth]{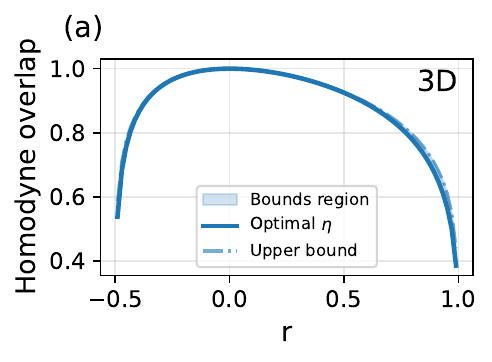}
\includegraphics[width=0.23\textwidth]{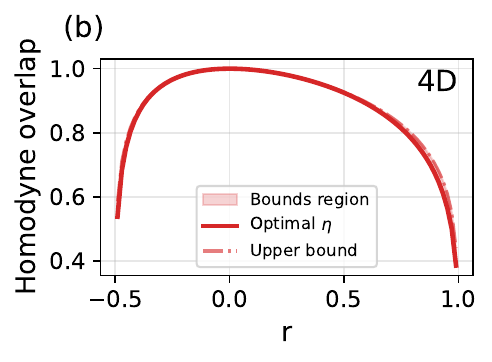} \\
\includegraphics[width=0.23\textwidth]{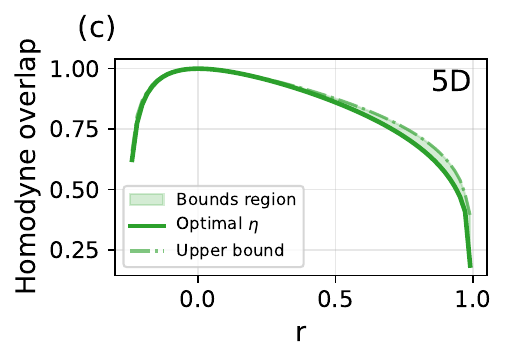}
\includegraphics[width=0.23\textwidth]{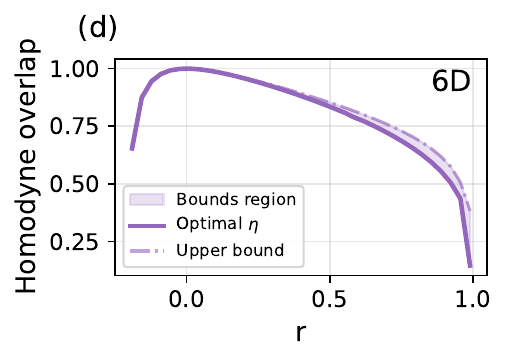}
\caption{\label{fig:bounds_eval_gaussian} Bound evaluation for Gaussian JSAs with $N\,\in\,\{3,4,5,6\}$ photons. Each panel shows the upper bound derived in Section \ref{sec:knapsack_bounds} (-.) and the numerically optimized overlap $\eta$ as a function of the correlation parameter $r$ (solid curve). }
\end{figure}

\begin{figure}[h]
\includegraphics[width=0.23\textwidth]{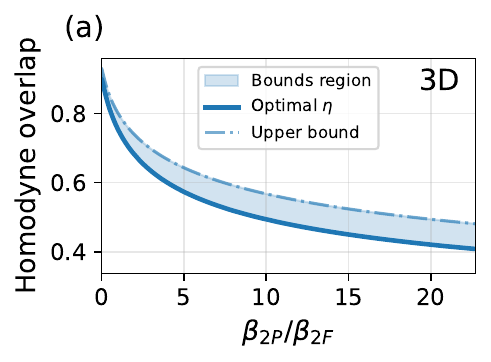}
\includegraphics[width=0.23\textwidth]{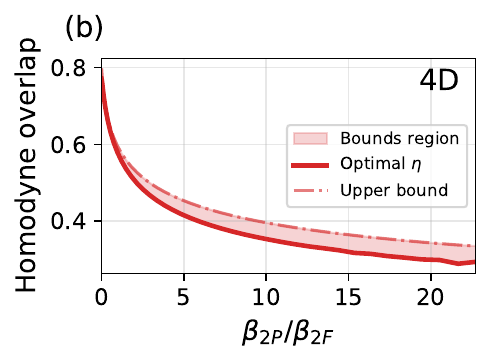}\\
\includegraphics[width=0.23\textwidth]{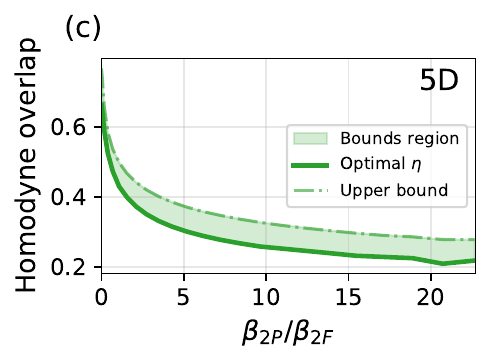}
\caption{\label{fig:bounds_eval_hopdc} Bound evaluation for JSAs associated with HOPDC with $N\,\in\,\{3,4,5\}$ photons. Each panel shows the upper bound derived in Section\;\ref{sec:knapsack_bounds} (-.), and the numerically optimized overlap $\eta$ as a function of the ratio $\beta_{2P}/\beta_{2F}$ (solid curve). The overlap $\eta$ remained closer to the bound for small ratios of $\beta_{2P}/\beta_{2F}$, with the gap widening as $\beta_{2P}\gg\beta_{2F}$ and the JSA becomes highly multimodal.} 
\end{figure}

We first evaluate how closely the homodyne overlaps obtained from the numerical optimization approach the theoretical bounds proposed in Section\;\ref{sec:knapsack_bounds}. Figures\;\ref{fig:bounds_eval_gaussian} and\;\ref{fig:bounds_eval_hopdc} show the upper bound, together with the numerically optimized homodyne overlap $\eta$ as a function of the correlation parameter $r$ for Gaussian JSAs, across photon-numbers $N\,\in\,\{3,4,5,6\}$, and for JSAs associated with HOPDC, we plot it as a function of the ratio $\beta_{2P}/\beta_{2F}$, across photon-numbers $N\,\in\,\{3,4,5\}$. The shaded region between the two bound curves highlights the interval within which $\eta$ is guaranteed to lie. 

In all cases, the optimized overlap remains below the predicted upper bound, confirming that the bound is valid across the whole range of correlation degrees and photon numbers considered. For Gaussian JSAs, the bound remains tight throughout. For JSAs associated with HOPDC, however, the optimal homodyne overlap $\eta$ follows the bound more closely at low correlation degrees ( $\beta_{2P}\ll \beta_{2F}$), and the gap widens as $\beta_{2P}$ increases in magnitude, which is representative of highly multimodal JSAs. This behaviour is likely related to the richer, non-Gaussian multimodal structure of HOPDC JSAs, compared to the Gaussian ones. 

The theoretical bound derived here thus provides a computationally inexpensive ceiling for estimating the maximum achievable visibility for both JSA families. In regimes where the bound is tight, namely for Gaussian JSAs and for HOPDC JSAs at low correlation degrees, it additionally serves as a good estimate of the actual achievable visibility. 

\subsubsection{\label{sec:lo_optim}
Optimal local oscillator distributions}

The optimal overlap $\eta$ quantifies the maximum achievable visibility, but the structure of the optimal local oscillator distribution carries additional information about the geometry of the JSA and the multimodal character of the photon-number state. In this section, we examine how the spectral distribution of the optimal LO varies with the degree of spectral correlations within the photon-number state and how it relates to the leading mode of the HOSVD decomposition. For Gaussian JSAs, the degree of correlations is determined by the correlation parameter $r$, as defined in Section\;\ref{sec:gaussian_jsa}, and for JSAs associated with HOPDC, the degree of correlations is determined by the ratio between the pump's and $N-$photon signal's group-velocity dispersions $\beta_{2P}/\beta_{2F}$, as defined in Section\;\ref{sec:hopdc_jsa}. 

\begin{figure}[ht]
\includegraphics[width=0.44\textwidth]{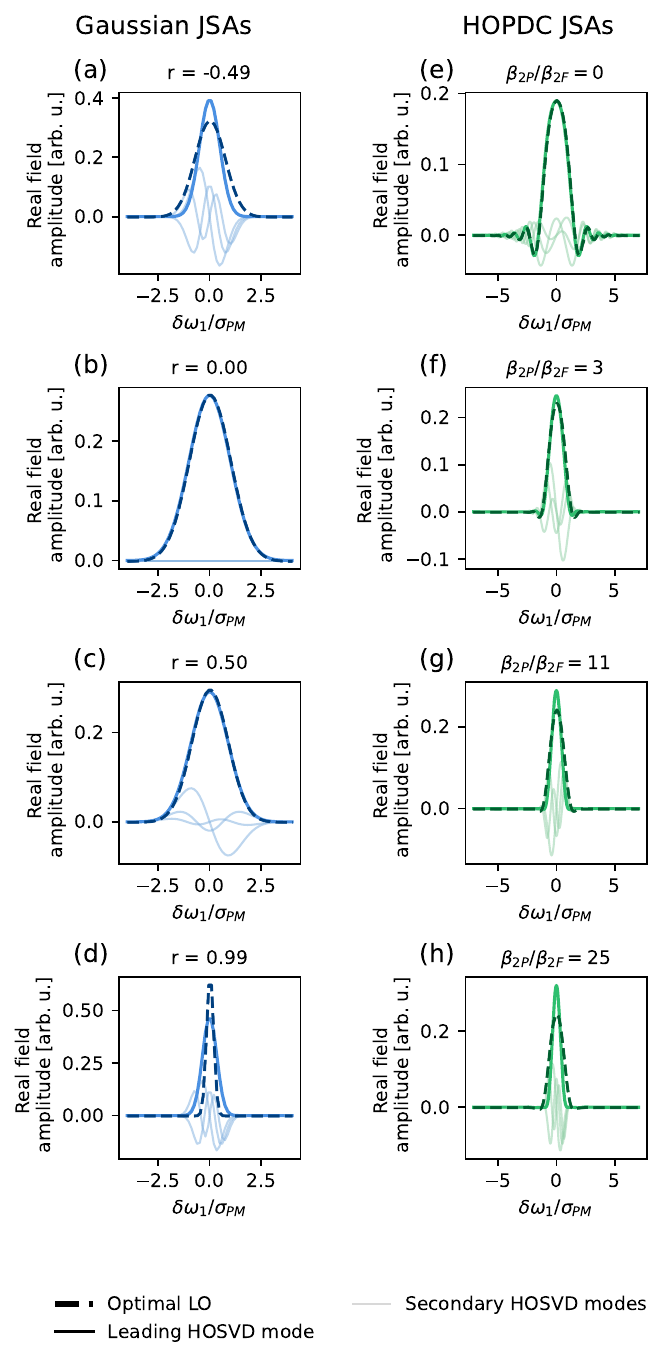} 
\caption{\label{fig:optimal_lo}Comparison between the optimal local oscillators and the HOSVD modal structure for different JSAs. In each panel, the dashed curve (--) represents the optimal local oscillator obtained from numerical optimization. The solid dark curve represents the leading HOSVD mode, while the lighter solid curves show the secondary HOSVD modes. The left column displays results for Gaussian JSAs (blue color), whereas the right column shows the corresponding distributions obtained from the JSAs associated with HOPDC (green color). The leading modes are normalized according to Eq. \eqref{eq:complete_orthonorm}.}
\end{figure}

In Figure\;\ref{fig:optimal_lo}, we show the optimal LO distributions for four representative correlation degrees, compared against the HOSVD decomposition for both families. In the left column we show the distributions for three-dimensional Gaussian JSAs ordered from top to bottom by increasing correlation parameter $r$, whereas in the left column, we show the distributions for three-dimensional JSAs associated with HOPDC, ordered from top to bottom by increasing ratio $\beta_{2P}/\beta_{2F}$. From direct inspection of the LO distributions, we observe that at low correlation degrees ($r\approx0$ or $\beta_{2P}\ll\beta_{2F}$), the optimal LO closely matches the leading HOSVD modes. This corresponds to low-rank JSAs or few-mode photon number states. As the degree of spectral correlations increases, contributions from secondary modes become visible and the alignment between the optimal LO and the leading HOSVD mode weakens. 

\begin{figure}[h]
\includegraphics[width=0.45\textwidth]{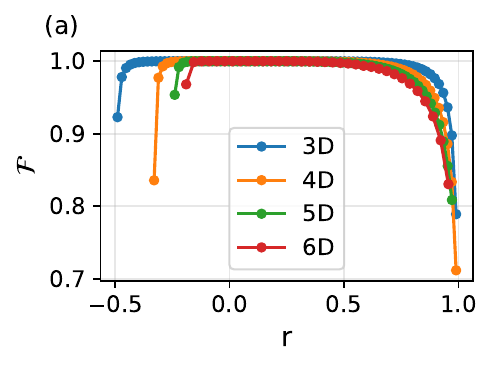}
\includegraphics[width=0.45\textwidth]{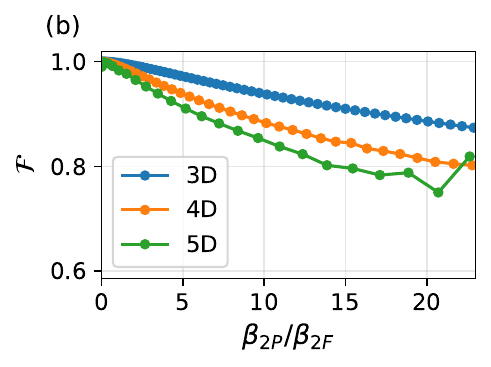}
\caption{\label{fig:overlap_lo} Fidelity optimal local oscillator distribution and leading HOSVD mode for (a) Gaussian JSAs and (b) JSAs associated with HOPDC.}
\end{figure}

To quantify the alignment between the optimal LO and the leading HOSVD mode, we define their fidelity as $\mathcal{F} = \left| \sum_{i} f_{i}\, u_{0i}^{\ast} \right|^{2}$, where $f_i$ and $u_{0i}$ denote the $i-$th components of the discrete optimal LO distribution and the leading HOSVD mode distribution, respectively. Figure \ref{fig:overlap_lo}(a) shows the fidelity as a function of the correlation parameter $r$, for Gaussian JSAs with $N\,\in\,\{3,4,5,6\}$. In Figure \ref{fig:overlap_lo}(b) we show an analogous analysis for JSAs associated with HOPDC, plotted as a function of the ratio $\beta_{2P}/\beta_{2F}$ for $N\,\in\,\{3,4,5\}$. 

\subsubsection{\label{sec:chirp} Special case: Complex JSAs associated to HOPDC with chirped pump pulses }

\begin{figure}[h]
\includegraphics[width=0.23\textwidth]{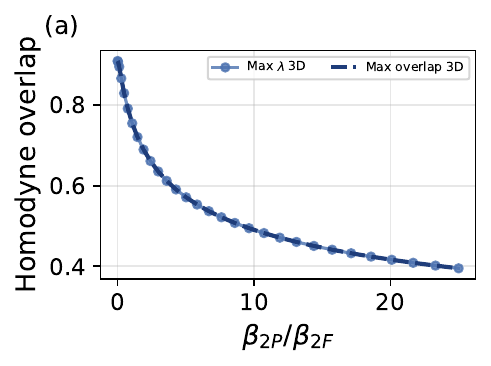}
\includegraphics[width=0.23\textwidth]{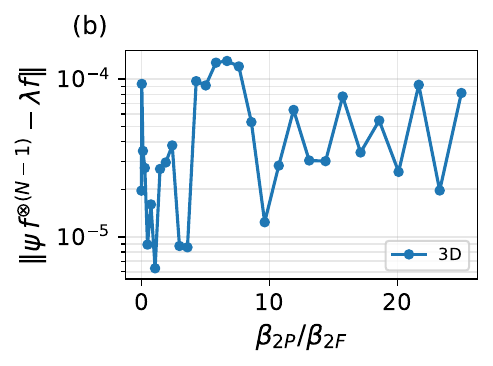}\\
\includegraphics[width=0.23\textwidth]{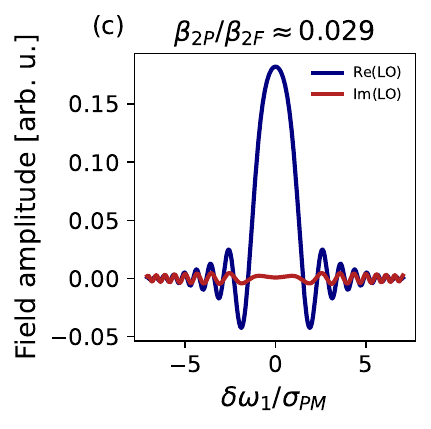}
\includegraphics[width=0.23\textwidth]{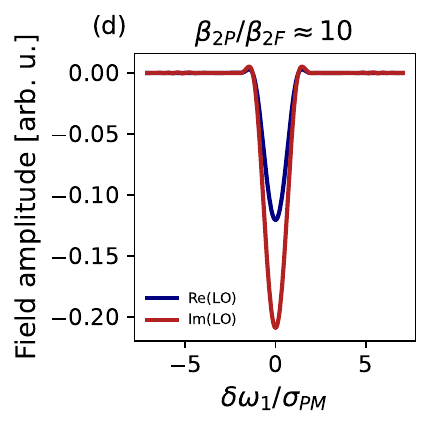}
\caption{\label{fig:eta_complex} Optimal homodyne overlap for a complex JSA associated with HOPDC and a chirped pump, as a function of the pump-to-signal GVD ratio $\beta_{2P}/\beta_{2F}$. (a) Optimal overlap $\eta^*$. (b) Residual of the eigenvalue condition at the converged solutions. (c,d) Real and imaginary parts of the optimal LO for two representative ratios, $\beta_{2P}/\beta_{2F}\approx0.03$ (c) and $\beta_{2P}/\beta_{2F}\approx10$ (d).}
\end{figure}

The examples considered so far have real-valued JSAs, for which the optimal LO is itself real up to a global phase. To test the methods presented here on a complex tensor, we consider HOPDC with a chirped pump envelope 
\begin{align}\label{eq:chirped_pump}
\alpha(\bm{\omega})=\alpha_0 \exp\left(-\frac{\left(\sum_i \left(\omega_i-\omega_{i0}\right)\right)^2}{2\sigma^2 (1+iC)}\right),
\end{align}
where $\alpha_0$ is a normalization constant, $\sigma$ is the bandwidth of the pump (See Eq. \eqref{sec:hopdc_jsa}), and $C$ is the chirp parameter. This quadratic chirp renders the JSA complex. 
We fix $C=2$, and $\sigma=\sigma_{PM}=\sqrt{4\pi/\ell|\beta_F|}$ (See Eq.\;\eqref{eq:pm_taylor}), and sweep across the pump to signal GVD ratio $\beta_{2P/\beta_{2F}}$ over the same range used in Section\;\ref{sec:numerical_eigen} for the unchirped case. Fig.\;\ref{fig:eta_complex}(a) shows the optimal overlap $\eta$ as a function of $\beta_{2P}/\beta_{2F}$. The curve is qualitatively similar to the unchirped result, indicating that the chirp does not alter the homodyne maximum achievable visibility. Fig.\;\ref{fig:eta_complex}(b) shows the residual of the eigenvalue condition $\lVert\bm{\psi} \cdot \bm{f}^{\otimes(N-1)}-\lambda\bm{f}\rVert_2$ evaluated at the converged solutions. Residuals remain in a precision range of $10^{-5}$ to $10^{-4}$ (relative to normalized quantities), confirming that the optimizer converges to a genuine eigenpair of the complex tensor. 
While the overlap itself was not affected by the pump chirp, the chirp makes the JSA complex, and therefore the optimal LO is complex as well. The structure of the optimal LO in this regime also depends strongly on the correlation regime. Figs.\;\ref{fig:eta_complex}(c,d) show the optimal LO for two sample JSAs with $\beta_{2P}/\beta_{2F}\approx0.03$ and $\beta_{2P}/\beta_{2F}\approx10$, respectively. For the weakly correlated case ($\beta_{2P}/\beta_{2F}\approx0.03$), the optimal LO is well approximated by a single dominant mode, consistent with the near-separable regime identified in Sec.\;\ref{sec:lo_optim}. 

\section{\label{sec:discussion}Discussion and outlook}

The main result of this work is a rigorous characterization of the optimal local oscillator for homodyne detection of photon number states in terms of the joint spectral amplitude tensor structure. We showed that the problem of maximizing the homodyne overlap reduces to finding the leading eigenpair of the JSA tensor, which is simultaneously equivalent to computing its spectral norm, solving the best rank-1 approximation problem, and evaluating the geometric measure of entanglement of the multiphoton state. 

For the computation of the optimal LO,  we proposed an efficient algorithm grounded in the geometry of the multidimensional JSA. In our proposal, the dominant rank-1 component that maximizes the overlap between the JSA and the local oscillator is obtained by using the Tucker HOSVD to initialize gradient-based optimization. Furthermore, considering the total symmetry of the photon number JSAs, we show that there is a direct relation between their HOSVD and the single-particle reduced density matrix: the Schmidt modes of the single-particle reduced density matrix coincide with the left singular vectors of the HOSVD factor matrix. This indicates that the HOSVD not only provides a principled initialization for the optimizer, but also contains meaningful spectral information about the multiphoton state. Building on this, we derived bounds on the optimal homodyne overlap based on the HOSVD procedure. These bounds are easy to evaluate, and serve as a diagnostic for the optimizer. The multilinear rank of the Tucker decomposition in this context serves as a measure of separability too: a rank-1 wavefunction is separable, and near-rank-1 wavefunctions admit a clear dominant mode that directly solves the optimal LO problem. This work shows that the Tucker HOSVD is a useful spectral analysis tool for high-dimensional quantum states, whose wavefunctions are intrinsically tensorial. This is complementary to other approaches, such as tensor networks and matrix product states.

We applied the proposed framework to Gaussian JSAs, across photon numbers $N\,\in\,\{3,4,5,6\}$, and those associated to higher-order parametric downconversion, across photon numbers $N\,\in\,\{3,4,5\}$, at varying correlation degrees. We found that for separable and near-separable states, the homodyne visibility is high and the optimal LO is well approximated by the leading HOSVD mode, which can be obtained without any nonlinear optimization. For highly multimode states with strong correlations, the maximum achievable visibility is strictly less than unity even at the global optimum, and the alignment between the optimal LO and the leading HOSVD mode weakens, making the nonlinear optimization step essential. 

Several directions for future work emerge from this work. The tensor eigenpair formulation developed here for homodyne detection can be extended, in principle, to other problems involving high-dimensional quantum states, including the optimization of sources and measurement schemes for higher-order squeezing and multipartite entanglement verification. On the computational side, a complementary direction is to exploit the symmetry of the JSAs or implement approaches such as Monte Carlo sampling to reduce the memory and computation runtime, enabling the method to extend to higher photon numbers. \\

\section*{Data availability statement}

The simulation data and algorithms supporting the findings of this study are available in the following repository:  \href{https://github.com/polyquantique/tensor_LO_optim}{https://github.com/polyquantique/tensor\_LO\_optim}.

\section*{Acknowledgements}
S. Molesky and P. Virally acknowledge financial support from the Natural Sciences and Engineering Research Council of Canada under Discovery Grant RGPIN-2023-05818 and the CGS-M program, and the Qu\'ebec Minist\`ere de l'\'Economie, de l'Innovation et de l'\'Energie (MEIE), as well as additional benefits provided from their affiliations to the Regroupement Qu\'eb\'ecois sur les Mat\'eriaux de Pointe, \doi{10.69777/309032}, the IVADO Research Consortium, and the Lassonde Deeptech Institute. N. Quesada and G.L. Osorio acknowledge financial support from the Natural Sciences and Engineering Research Council of Canada as well as additional benefits provided from their affiliations to  	L'Institut transdisciplinaire d'information quantique (INTRIQ), \doi{10.69777/340940}.

\bibliography{bib.bib}

\clearpage

\appendix

\section{\label{sec:comp_perform}Computational performance and scalability}

\subsection{Comparison with trust-region basin-hopping on a representative case}
To assess the relative computational cost of our optimization strategy, we compared it against Trust-Region Basin-Hopping (TRBH) on a Gaussian JSA with correlation coefficient $r=0.96$ (highly-multimode case), for $N=3,4,5$. Both methods used the same local solver (L-BFGS-B) with stopping tolerances (objective function tolerance and gradient tolerance $f_{\text{tol}}=g_{\text{tol}}=10^{-8}$), and each outer iteration was allotted up to 1000 inner iterations, with early stopping upon convergence. We report the number of outer iterations for TRBH and our method (adaptive penalty-homotopy method, APH) required to reach the verified global optimum:
\begin{table}[h]
\centering
\caption{Comparison between our adaptive penalty-homotopy strategy (APH) and Trust-Region Basin-Hopping (TRBH) for a Gaussian JSA with correlation coefficient $r=0.96$.}
\label{tab:trbh_comparison}
\begin{tabular}{c c c c c c}
\hline\hline
\makecell{$N$} &
\makecell{grid size \\ $n$} &
\makecell{tensor size \\ $n^N$} &
\makecell{TRBH outer \\ iters.} &
\makecell{APH outer \\ iters.} &
\makecell{Ratio \\ outer iters.} \\
\hline
\hline
3 & 60 & $2.16\times10^{5}$ & 250 & 24 & 10.4 \\
4 & 60 & $1.30\times10^{7}$ & 249 & 28 & 8.9  \\
5 & 30 & $2.43\times10^{7}$ & 306 & 30 & 10.2 \\
\hline\hline
\end{tabular}
\end{table}

In all three cases, both methods converged to the same maximum objective up to $10^{-6}$ precision and the same local oscillator amplitude, confirming that the reported optimum is global. Our method consistently required an order of magnitude fewer outer iterations than TRBH to reach it.
Note that the grid size $n$ was reduced from 60 to 30 at $N=5$: the JSA tensor is stored explicitly with $n^N$ entries, so memory requirements grow exponentially in $N$ for fixed $n$; $n=60$ in $N=5$ exceeded the available computing memory. Notice that runtime cost depends on $n$ and $N$, so the outer-iterations presented here should not be read as direct runtime comparison, and are only comparable in optimizations with the same grid size $n$ and dimensionality $N$.

\begin{figure}[h]
\centering
\includegraphics[width=\columnwidth]{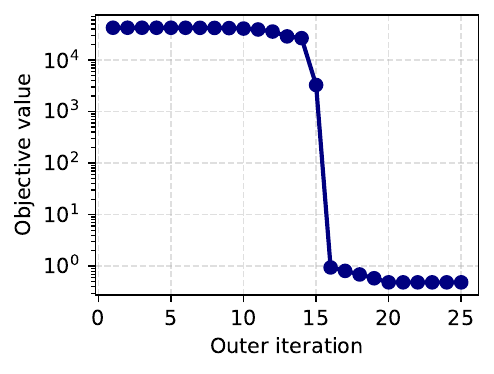}
\caption{Representative convergence trace of the homodyne overlap $\eta$ as a function of outer iteration for our method (Gaussian JSA, $r=0.96$, $N=4$). During the initial outer iterations, the constraint is relaxed via the $2N$-norm homotopy, so $\|\bm{f}\|_2 \neq 1$ and the raw value of $\eta$ is not physically meaningful. As the homotopy directs the constraint toward $\|\bm{f}\|_2=1$, $\eta$ drops sharply around iteration~15--16 and stabilizes at its converged value, once the unit-norm constraint is satisfied.}
\label{fig:convergence_eta}
\end{figure}

Fig. \ref{fig:convergence_eta} shows a representative convergence trace of the objective $\eta$ over outer iterations for our method. During the initial outer iterations, $\eta$ takes large values because the iterate is not yet constrained to the unit sphere. As the norm homotopy directs the constraint toward the 2-norm, $\eta$ drops and stabilizes at its final value once the constraint is satisfied.

\subsection{Practical limitations}
The main scalability bottleneck is the explicit storage of the JSA tensor, which scales as $O(n^N)$; this already forced a reduction in grid resolution for JSAs with $N\geq5$ and limited us to $N\leq 6$ within available memory. Per-iteration cost scales similarly, since objective and gradient evaluations require contracting the full tensor. Extending the method to higher photon numbers will require reducing the JSA storage. Two strategies that could be implemented for that task are  exploiting the JSA symmetry or implementing Monte Carlo sampling to reduce the memory and computation runtime, enabling the method to extend to higher photon numbers.

\end{document}